\input mnrass.sty
\pageoffset{-2.5pc}{0pc}

 

\Autonumber  


\pagerange{000--000}
\pubyear{1996}
\volume{000}

\begintopmatter  

\title{Stellar models for very low mass main sequence stars: the role
of model atmospheres. }

\author{E. Brocato$^{1,2}$, S. Cassisi$^{1,3}$ \& V. Castellani$^{2,4}$}

\affiliation{$^1$Osservatorio Astronomico di Collurania, Via M. Maggini,
 I-64100, Teramo, Italy - E-Mail: brocato,cassisi@astrte.te.astro.it}
\affiliation{$^2$Istituto Nazionale di Fisica Nucleare, LNGS, 67010 Assergi, L'Aquila, Italy}
\affiliation{$^3$Universit\'a degli studi de L'Aquila, Dipartimento di Fisica,
Via Vetoio, I-67100, L'Aquila, Italy}
\affiliation{$^4$Dipartimento di Fisica, Universit\'a di Pisa, Piazza Torricelli 2, 
I-56100, Pisa, Italy- E-Mail: vittorio@astr1pi.difi.unipi.it}

\shortauthor{Brocato, Cassisi \& Castellani}
\shorttitle{Stellar models for VLM stars.}



\abstract

\tx
We present Very Low Mass stellar models as computed including non-grey 
model atmospheres  for selected assumptions about the star metallicities.
The role of atmospheres is discussed and the models are compared 
with  models based on the Eddington approximation and
with similar models appeared in the recent literature.
Theoretical predictions concerning both the HR diagram location
and the mass-luminosity relation are presented and discussed in terms of
expectations in selected photometric bands.
Comparison with available observational data concerning
both galactic globular clusters and dwarfs in the solar
neighborhood reveals a satisfactory agreement together with
the existence of some residual mismatches.  


\maketitle  

\section{Introduction}

\tx 
The long-standing theoretical interest in Very Low Mass (VLM) Main Sequence
stars has been recently rejuvenated according to the increasing amount 
of VLM objects observed with the Hubble
Space Telescope as well as thanks to CCD parallax
determinations (Monet et al. 1992, Dahn et al. 1995, Tinney 1996) which are
increasing the amount of absolute magnitudes available for nearby dwarfs.
This interest is further enhanced by the suggestion that an appreciable 
fraction of the baryonic mass in most galaxies
could be in the form of VLM stars and Brown Dwarfs. 
To have light on such an observational scenario one needs reliable theoretical 
predictions about VLM stellar structures and, in particular,
accurate mass-luminosity relations allowing the evaluation of reliable 
mass functions and, in turn, reliable estimates of the mass 
density of VLM stars in the Galaxy.

\par
The temperature scale of M dwarfs has been for 
long time an unsettled problem and
it is still a key ingredient for understanding the location of VLM stars
in the HR diagram (Bessel 1995, Leggett et al. 1996).
Since the pioneering works  by Limber (1958), Hayashi 
\& Nakano (1963) and Ezer \& Cameron (1967), numerous investigators
(D'Antona 1987; VandenBerg et al. 1988; Burrows, Hubbard \& Lunine 1989;
Dorman, Nelson \& Chau 1989; Burrows et al. 1993; D'Antona \& Mazzitelli 1994, 
1996; Baraffe et al. 1995; Baraffe \& Chabrier 1995; Baraffe \& Chabrier 1996;
Alexander et al. 1997; Kroupa \& Tout 1997) have already devoted their attention to the 
structural properties and the evolutionary behavior of VLM stars. 
As early recognized, one knows that convection in VLM stars is very efficient 
throughout all the structure and that a VLM structure is very 
nearly adiabatic. As a consequence,  theoretical predictions for
the interior of such peculiar objects do not depend on radiative opacities, 
nor on the choice of the mixing length parameter governing the superadiabatic
convection.

\par
In spite of these advantages, VLM stellar models critically 
depend on the evaluations of both opacity and equation 
of state (EOS) for a low temperature, high density gas,
where molecules, grains and non-ideal gas effects play a relevant role.
Concerning radiative opacities, large efforts have been recently 
devoted to properly include in opacity computations the 
effects of molecules and grains, with particular emphasis on the
treatment of the opacity due to the $\rm H_2O$ and $\rm TiO$ molecules 
(see, e.g., Dorman et al. 1989, Allard \& Hauschildt 1994, 1995a, 1995b, 
Bessel 1995). Low temperature opacity tables presented by Alexander \& 
Ferguson (1994) and  Alexander (1994) represent a significant improvement 
of previous evaluations  given by Cox \& Tabor (1976), 
Alexander, Johnson \& Rypma (1983), and  Kurucz (1992).

\par
Substantial progresses (Saumon 1994) in the EOS for VLM stars are due 
to Saumon, Chabrier \& Van Horn (1995) (but see also Saumon \& Chabrier 1992), 
who provided an EOS for dense and cool matter based on a 
detailed description of the physics at work. However, the theory of 
VLM structures requires another critical ingredient,
i.e.,  the boundary conditions for the inner stellar 
structure, as given by a suitable treatment of stellar atmospheres.
Grey model atmospheres usually assume that the 
optically thin region lying above the photosphere is not affected by 
convection.  However, a similar assumption
is not well-grounded for several VLM atmospheres, since atmospheric 
opacities can be very large due to the presence of molecules 
whereas the adiabatic gradient is small due 
to the dissociation of $\rm H_2$. As a consequence the atmosphere can 
be unstable
against convection also at very small optical depth ($\tau$).
Moreover, non-grey effects in the stellar atmospheres are often 
neglected, though, theoretical evidences (Saumon et al. 1994) 
suggest that such effects could play a significant role in governing 
the atmospheric structure.

\par
In the last decade, significant improvements in model atmosphere 
for late  M dwarfs have been presented  
by Allard (1990), Kui (1991), Brett \& Plez (1993), Saumon et al. (1994), 
Allard \& Hauschildt (1995) and Brett (1995a, 1995b). 
Similar models, and in particular the models provided by
Saumon et al. (1994) for zero metallicity mixtures and by Allard \& 
Hauschildt (1995) and Brett (1995a,1995b) for finite metallicities, 
represent a substantial progress of our knowledge in the field.
Nevertheless, Bessell (1995) has  pointed out the 
still existing differences between different models, as due
to the different evaluations of the contribution to opacity of
$\rm H_2O$ and TiO molecules
and to the different opacity averaging technique adopted in the various works.

\par
From an evolutionary point of view, the need for accurate model atmospheres
in computing VLM stellar models has been firstly stressed by 
Burrows et al. (1993) and more recently reinforced  by Baraffe et al. (1995),
Baraffe \& Chabrier (1996), Chabrier, Baraffe \& Plez (1996),
M\'era, Chabrier \& Baraffe (1996).
In previous works (Alexander et al. 1997, hereinafter Paper I;
Brocato et al. 1997a), we have presented a theoretical approach 
to the evolutionary behavior of VLM stars, showing that the adoption 
of the most updated  equation of state (Saumon et al. 1995) and low temperature 
opacity (Alexander \& Ferguson 1994), but still relying on an 
approximate treatment of the stellar atmosphere,
allows a rather satisfactory agreement between observation and 
theoretical predictions for the Color-Magnitude (CM)  diagrams and the 
mass-luminosity relation
of both metal poor and solar metallicity VLM objects.
We thus concluded Paper I with the statement that {\sl "
the use of a $T(\tau)$ relation in computing stellar models 
has to be regarded as a first order but not-too-bad approximation 
to the expected evolutionary behavior"}.
In this paper  we will go deeper in that matter, discussing
VLM stellar models including updated outer boundary conditions 
for various selected assumptions about the star metallicities.

\par
The layout of this paper is as follows. Next section will provide 
some general informations about the models, with particular  
attention to the adopted grid of  model atmospheres. In \S 3, 
models computed adopting model atmospheres will be compared 
with similar models based on the Eddington approximation and
with stellar models already appeared in the literature. Section 4 
presents the comparison between observational data and theoretical 
results. Conclusions will follow in \S 5.
\section{Boundary conditions.}
\tx

Models presented in this paper adopt the same EOS and the same
opacity evaluations as in Paper I. The main difference 
with models in Paper I  is the different approach adopted for 
deriving the outer boundary conditions, i.e.,  temperature and  
pressure of the gas at the basis of the atmosphere.
In Paper I the atmospheric integrations were performed adopting
the Krishna-Swamy (1966) solar scaled $T(\tau)$ formula 
until reaching the optical depth $\tau=2/3$ or, alternatively, until the 
onset of convection, where the standard
mixing length theory was used to evaluate the degree of superadiabaticity.
In the present paper outer boundary conditions will be evaluated 
by adopting suitable non-grey model atmospheres.

\par
To our knowledge, the most up\-dated model atmosphe\-res presently
available are the ones computed by Brett (1995 a,b, hereinafter B95) 
and the "next generation"  of the Allard \& Hauschildt 
models (Allard \& Hauschildt 1997, hereinafter NG97).
Both sets of models include updated (but not identical)
line lists and in both sets the numerous 
atomic and molecular opacity sources
have been modeled with the accurate opacity sampling
technique (for a detailed discussion on this point see also Bessell 1995).
At the time when present computations have been performed NG97 model atmospheres
were under-computing (Leggett et al. 1996,
Chabrier et al. 1996) and results were available for solar metallicity only. 
B95 models are actually available from solar metallicity down to
metallicities as low as $[M/H]\sim-2.0$. In a recent paper 
Chabrier et al. (1996, see also Baraffe \& Chabrier 1996) have discussed the 
comparison between stellar models computed alternatively adopting
boundary conditions from NG97, from B95, or from "old" ({\sl Base}) model
atmospheres by Allard \& Hauschildt (1995, hereinafter AH95), emphasizing 
as a final result {\sl "the excellent agreement between the 
observation and the stellar models based on B95 and NG97, 
while models based on AH95 clearly underestimate the flux
in the V band"}. As pointed out by these authors, 
this occurrence has to be regarded as a plain evidence of the accuracy 
recently achieved in computing model atmospheres for M
dwarfs.

\par
In order to investigate VLM stellar structures in this paper 
we will adopt B95 atmospheric models, which cover the ranges 
$4000K\ge{T_{eff}}\ge2600K$, with $\log{g}=4.5, 5.0$ and for metallicities
$Z= 0.0002, 0.002, 0.02, 0.04$, adopting for the mixing length 
$l=1.5 H_P$, where $H_P$ is the pressure scale height. However, low metallicity VLM models 
reach larger effective
temperature (see, e.g.,  Paper I) and the "critical" temperature ($T_{eff}^{crit}$)
defining the lower limit for the validity of the Eddington approximation
also increases above $4000K$ (see Baraffe et al. 1995 
and reference therein). Therefore, at the lower metallicities B95 models 
will be implemented  with  Kurucz's (1993) model atmospheres which
are available for effective temperature $T_{eff}\ge3500K$
and in a large range of gravity.
One has to notice that Kurucz models 
lack the important $\rm H_2O$  opacity contribution
which can be of major relevance in cool VLM objects. 
However, Brett (1995a; but see also Bessel 1995)
has already found that, regardless of the missing
$\rm H_2O$ opacity and the different mixing length parameter 
(Kurucz adopts $l=1.25 H_P$),
at $4000K$ the two model atmospheres are almost identical, 
whereas at $T_{eff}=3500K$ significant differences appear.
Thus Kurucz models can be safely used only for temperatures above
$4000K$. 

\par
When using model atmospheres as boundary conditions one has to remind that
in stellar interiors the radiative flux is estimated by means of 
the diffusion equation 
which is the limit of the transfer equation for large optical depth 
(see Mihalas 1978 for a detailed discussion on this
topic). One should thus pay attention that the assumed limit 
between the internal 
structure (where the diffusive approximation is used) and
the outer atmosphere (where this approximation is no longer valid) 
is  located at a Rosseland optical depth $\tau_{Ros}$ large enough 
for the diffusion approximation to be valid. Morel et al. 
(1994) have recently shown that for solar abundances
a suitable value for $\tau_{Ros}$ is about 10. In the ATLAS9 code 
(Kurucz 1993), the monochromatic flux is computed through
the diffusion approximation when $\tau_\nu$ 
(i.e. the optical depth at the frequency $\nu$) is larger or equal to 20.
Thus a safe procedure is to connect
the ATLAS9 model atmosphere with a model of stellar interior at 
an optical depth insuring the condition $\tau_{\nu}\ge20$ for 
all frequencies, that it means $\tau_{Ros}>20$.
According to the evidence that  physical mechanisms related to the 
energy transport in the outer layer are accounted for in a more 
realistic treatment in the atmosphere codes and also to fulfill the
quoted condition, we fixed the bottom of the atmosphere 
at $\tau_{Ros}=100$. The same value for the boundary  limit has been 
adopted also by Baraffe \& Chabrier (1996) and this occurrence 
should allow a safe comparison between our models
and the ones provided by the quoted authors.
\figure{1}{S}{80mm}{\bf Figure 1. \rm The location in the HR diagram of 
solar metallicity, VLM stellar models computed under various assumptions 
(as labeled) for the boundary condition. 
Dotted lines connect stellar models with the same mass.}

\par
Nevertheless, it appears interesting to investigate how far
the evolutionary behavior and in particular the location of
the models in the HR diagram could be affect by different choices 
concerning the outer boundary limit.
For this aim stellar models with solar metallicity have been computed 
under boundary conditions taken at different values of the Rosseland 
optical depth.

\par
The computed stellar models cover the mass range 
$0.095M_\odot\le{M}\le0.6M_\odot$, assuming an original Helium abundance
as given by Y=0.27.
To be consistent with evolutionary computations shown in Paper I,
we have adopted for the mixing length $l=2.2 H_P$. The adopted value is
different than the values adopted by Brett (1995) and Kurucz (1993)
in computing their model atmospheres.
However, it has been verified performing additional numerical tests 
that reasonable variations on the value of $l$ have no effects at all 
on the determination of the stellar radius of stars less massive than about $0.5M_{\odot}$
(see Paper I and references therein), and only minor changes for stars
with mass around $0.6M_{\odot}$. Since in present work, we are mainly dealing with stellar
models with mass $M\le0.6M_\odot$, the results are not affected by the choice on the mixing 
length parameter.
Moreover, as far as concerns the model atmosphere computations, Brett (1995) has 
clearly shown that a variation of the mixing length parameter in the
range 1.0 to 2.0 produces only minor effects at effective temperatures
around $4000K$, and that these structural effects
decreases rapidly decreasing the effective temperature of the models.
\figure{2}{S}{80mm}{\bf Figure 2. \rm The HR diagram location of 10Gyr old models
for the chemical composition: Z=0.002 - Y=0.23, computed under different assumptions
about the outer boundary conditions.}

\par
Figure 1 show the HR diagram location of  10Gyr old stellar models computed
adopting $\tau_{Ros}$: 0.1, 1.0, 100. In all models
the outer boundary conditions have been obtained from B95 model 
atmospheres alone. For the aim of comparison, the $T({\tau})$ models 
as given in  Paper I are also presented. As expected,  $T(\tau)$ models 
are systematically hotter and brighter than B95 models with the same mass. 
The maximum departure occurs at ${\log}T_{eff}\approx3.52$ which 
corresponds to $M\approx0.30M_{\odot}$, with differences
$\Delta{T_{eff}}\approx150 K$ and $\Delta{\log}L/L_\odot\approx0.08$.
The origin of such behavior is related to both the $\rm H_2$ 
dissociation mechanism and to the resulting penetration of the convection 
in the optical thin outer atmosphere (Auman 1969; Dorman et al. 1989; 
Burrows et al. 1993). 

\par
Again as expected, one finds that at smaller values of  ${\tau}$ the 
location of the models is 
largely sensitive to the adopted value of the Rosseland optical depth 
where outer boundary conditions are fixed. In fact by changing
$\tau_{Ros}$ from 0.1 to 1.0 the models become cooler by $\approx (70-80)K$ 
and fainter by $\Delta{\log}L/L_{\odot}\approx0.05-0.1$. 
However, for $\tau_{Ros}\ge1.0$ the models are not significantly affected 
by changes in the adopted value of $\tau_{Ros}$. As a result,
one finds the evidence that the evolutionary results are not dependent
on the choice about the point where the atmosphere is matched 
with the envelope, once that a sufficiently large value ($\tau_{Ros}>20-30$) 
is adopted in the evolutionary codes.

\par
Bearing in mind such a scenario, stellar models for lower metallicities
have been computed under the above quoted assumptions, for Z=0.002 and 
0.0002 (as in B95), assuming now Y=0.23 and adopting outer boundary conditions 
either from B95 or from Kurucz's model. For each given
metallicity, a set of stellar models still relying on the Eddington
approximation has been also computed in order to investigate the effect of
incorporating model atmospheres in stellar computations as well as
to obtain indication on the value of the critical temperature $T_{eff}^{crit}$.
Figures 2 and 3 show the results of these computations.
As expected, even a quick inspection of these figures shows  that at 
effective temperatures around $4000K$ a good match is achieved between the 
different models, supporting the extension of B95 with Kurucz models. 
More in details, when Z=0.002 one may notice in Figure 2 
the fine smooth transition 
between the $T(\tau)$ stellar models and the 
K93 ones which occurs at $T_{eff}\approx4400K$ for 
a $\approx 0.55M_{\odot}$ model. The same smooth transition occurs
between  K93  and B95 models  at $T_{eff}\approx4000K$
for a $\simeq0.42M_{\odot}$ object. This finding appears in fine 
agreement with the results by Brett (1995a, his figure 8) when comparing 
his model atmospheres with the ATLAS9 results.
The evidence that for $T_{eff}<4000K$ the K93 model sequence 
"converges" towards the $T(\tau)$ models is simply due to the missing 
$\rm H_2O$ opacity in the Kurucz's (1993) model atmospheres.
The stellar models computed by adopting the Eddington 
approximation are hotter and brighter
than B95 models. The larger discrepancy is present at $T_{eff}\approx3800K$ 
where it is $\approx140K$ in effective temperature and
$\Delta{\log}L/L_{\odot}\approx0.055$ in luminosity. 
\figure{3}{S}{80mm}{\bf Figure 3. \rm As in Figure 2, but for Z=0.0002 - Y=0.23.}

The HR diagram location of the stellar models with metallicity Z=0.0002
is displayed in Figure 3. In this case the match between the 
$T(\tau)$ and  K93 models is achieved at 
$T_{eff}\approx5000K$. This behavior appears in good agreement with 
result by Burrows et al (1993) concerning the increase of $T^{crit}_{eff}$ 
when the heavy elements abundance is decreased. As expected, 
even at this very low metallicity  B95 models agree with K93 ones
at $T_{eff}\simeq4000K$. The maximum discrepancy between $T(\tau)$ 
models and stellar structures computed by adopting
accurate model atmospheres appears at $T_{eff}\approx(4100-4200)K$, 
where one finds differences of the order
of $\approx180K$ in $T_{eff}$ and $\approx0.003$ in ${L/L_{\odot}}$
\section{Stellar models.}
\tx

For each given metallicity, we selected as the {\sl "best"} sequence of models
the one obtained by adopting in the proper range of validity 
the $T(\tau)$, the K93 and the B95 stellar models, paying attention that 
a fine and smooth match is obtained between the models computed 
under different assumptions about the outer boundary conditions.
Figure 4 displays the run of present {\sl "best"} MS in the HR diagram for
the labeled assumptions about metallicity.
Tables 1 through 3 give the luminosity, the effective temperature, 
the absolute visual magnitude and predicted colors (in the 
standard Johnson-Cousins system for $VRI$ and the CIT system for $K$) for  
10Gyr old {\sl "best"} models for the various selected metallicities. 
Magnitudes and colors  
have been evaluated by adopting bolometric correction and color 
temperature relation from Kurucz (1993)  
implemented at effective temperatures lower than $4000K$ with 
similar evaluations given by Allard \& Hauschildt (1995) or, for 
solar metallicity, by Allard et al. 1997 (Allard 1996), 
both properly shifted to overlap at each metallicity Kurucz's evaluation at 
the fitting  point $T_{eff}=4000K$

\figure{4}{S}{80mm}{\bf Figure 4. \rm The HR diagram location of 10Gyr old models
for the labeled assumptions on the chemical composition.}
\table{1}{S}{\bf Table 1. \rm Mass, luminosity, effective temperature, absolute visual magnitude
and colors for stellar models with solar metallicity and Y=0.27, at age equal to 10Gyr.}
{\halign{%
\rm#\hfil& \hskip3pt\hfil\rm#\hfil&\hskip4pt\hfil\rm#\hfil&\hskip7pt\hfil\rm#\hfil&\hskip4pt\hfil\rm#\hfil& \hskip4pt\hfil\rm\hfil#&\hskip4pt\hfil\rm\hfil#\cr
$M/M_\odot$ & $\log{L/L_{\odot}}$ & $\log{T_e}$ & $M_V$ & $(V-I)$ & $(V-R)$ & $(V-K)$ \cr 
\noalign{\vskip 10pt}
       .600 &  -1.152 &   3.590 &   8.954 &   1.965 &   1.057 &   3.817 \cr  
       .550 &  -1.313 &   3.570 &   9.656 &   2.192 &   1.147 &   4.180 \cr  
       .500 &  -1.468 &   3.554 &  10.276 &   2.352 &   1.220 &   4.450 \cr  
       .450 &  -1.616 &   3.541 &  10.806 &   2.455 &   1.273 &   4.635 \cr  
       .400 &  -1.751 &   3.531 &  11.263 &   2.531 &   1.311 &   4.771 \cr 
       .350 &  -1.881 &   3.524 &  11.674 &   2.587 &   1.339 &   4.866 \cr  
       .300 &  -2.002 &   3.518 &  12.056 &   2.639 &   1.365 &   4.953 \cr 
       .280 &  -2.061 &   3.516 &  12.230 &   2.657 &   1.373 &   4.983 \cr  
       .250 &  -2.159 &   3.511 &  12.553 &   2.707 &   1.397 &   5.066 \cr 
       .200 &  -2.360 &   3.501 &  13.254 &   2.841 &   1.457 &   5.285 \cr 
       .180 &  -2.459 &   3.495 &  13.652 &   2.944 &   1.502 &   5.452 \cr 
       .150 &  -2.638 &   3.482 &  14.420 &   3.152 &   1.600 &   5.809 \cr
       .120 &  -2.887 &   3.459 &  15.721 &   3.571 &   1.813 &   6.557 \cr  
       .100 &  -3.147 &   3.429 &  17.462 &   4.213 &   2.175 &   7.737 \cr 
       .098 &  -3.181 &   3.425 &  17.704 &   4.301 &   2.229 &   7.905 \cr 
       .097 &  -3.200 &   3.422 &  17.868 &   4.365 &   2.268 &   8.029 \cr 
       .095 &  -3.239 &   3.419 &  18.084 &   4.431 &   2.309 &   8.155 \cr}} 
\table{2}{S}{\bf Table 2. \rm As in Table 1, but for metallicity Z=0.002 and Y=0.23.}
{\halign{%
\rm#\hfil& \hskip3pt\hfil\rm#\hfil&\hskip4pt\hfil\rm#\hfil&\hskip7pt\hfil\rm#\hfil&\hskip4pt\hfil\rm#\hfil& \hskip4pt\hfil\rm\hfil#&\hskip4pt\hfil\rm\hfil#\cr
$M/M_\odot$ & $\log{L/L_{\odot}}$ & $\log{T_e}$ & $M_V$ & $(V-I)$ & $(V-R)$ & $(V-K)$ \cr 
\noalign{\vskip 10pt}
       .700 &   -.514 &   3.717 &   6.273 &    .952 &    .540 &   1.929\cr  
       .600 &   -.833 &   3.676 &   7.240 &   1.155 &    .661 &   2.381\cr 
       .550 &  -1.030 &   3.643 &   7.945 &   1.379 &    .800 &   2.788\cr  
       .500 &  -1.226 &   3.617 &   8.683 &   1.617 &    .923 &   3.158\cr   
       .450 &  -1.407 &   3.600 &   9.292 &   1.764 &    .997 &   3.381\cr    
       .420 &  -1.507 &   3.592 &   9.571 &   1.797 &   1.013 &   3.431\cr 
       .400 &  -1.568 &   3.588 &   9.739 &   1.814 &   1.021 &   3.458\cr  
       .380 &  -1.624 &   3.584 &   9.901 &   1.835 &   1.029 &   3.490\cr   
       .350 &  -1.704 &   3.580 &  10.121 &   1.855 &   1.037 &   3.522\cr  
       .320 &  -1.773 &   3.577 &  10.312 &   1.872 &   1.043 &   3.549\cr   
       .300 &  -1.826 &   3.574 &  10.466 &   1.891 &   1.048 &   3.579\cr   
       .250 &  -1.981 &   3.567 &  10.910 &   1.941 &   1.060 &   3.657\cr    
       .200 &  -2.192 &   3.556 &  11.553 &   2.037 &   1.075 &   3.806\cr   
       .180 &  -2.297 &   3.549 &  11.888 &   2.097 &   1.085 &   3.900\cr   
       .150 &  -2.486 &   3.536 &  12.522 &   2.229 &   1.111 &   4.101\cr   
       .120 &  -2.752 &   3.511 &  13.546 &   2.516 &   1.177 &   4.533\cr   
       .110 &  -2.879 &   3.498 &  14.098 &   2.699 &   1.230 &   4.797\cr   
       .100 &  -3.066 &   3.474 &  15.087 &   3.092 &   1.373 &   5.370\cr   
       .099 &  -3.092 &   3.470 &  15.262 &   3.167 &   1.405 &   5.499\cr  
       .098 &  -3.119 &   3.466 &  15.439 &   3.240 &   1.437 &   5.626\cr   
       .096 &  -3.182 &   3.456 &  15.865 &   3.429 &   1.525 &   5.921\cr   
       .095 &  -3.216 &   3.451 &  16.081 &   3.523 &   1.572 &   6.059\cr   
       .093 &  -3.294 &   3.438 &  16.651 &   3.793 &   1.710 &   6.450\cr   
       .092 &  -3.338 &   3.430 &  16.991 &   3.959 &   1.799 &   6.683\cr}}   
\table{3}{S}{\bf Table 3. \rm As in Table 1, but for metallicity Z=0.0002 and Y=0.23.}
{\halign{%
\rm#\hfil& \hskip3pt\hfil\rm#\hfil&\hskip4pt\hfil\rm#\hfil&\hskip7pt\hfil\rm#\hfil&\hskip4pt\hfil\rm#\hfil& \hskip4pt\hfil\rm\hfil#&\hskip4pt\hfil\rm\hfil#\cr
$M/M_\odot$ & $\log{L/L_{\odot}}$ & $\log{T_e}$ & $M_V$ & $(V-I)$ & $(V-R)$ & $(V-K)$ \cr 
\noalign{\vskip 10pt}
       .700 &   -.385 &   3.758 &   5.910  &   .805  &   .451  &  1.580\cr  
       .600 &   -.743 &   3.712 &   6.883  &   .988  &   .553  &  1.966\cr    
       .500 &  -1.128 &   3.653 &   8.082  &  1.289  &   .729  &  2.564\cr  
       .450 &  -1.311 &   3.634 &   8.661  &  1.417  &   .807  &  2.785\cr    
       .400 &  -1.463 &   3.622 &   9.132  &  1.506  &   .861  &  2.936\cr   
       .350 &  -1.581 &   3.616 &   9.473  &  1.550  &   .890  &  3.010\cr  
       .300 &  -1.702 &   3.611 &   9.809  &  1.583  &   .913  &  3.067\cr   
       .250 &  -1.849 &   3.605 &  10.222  &  1.627  &   .940  &  3.138\cr   
       .200 &  -2.056 &   3.595 &  10.813  &  1.698  &   .977  &  3.233\cr  
       .180 &  -2.161 &   3.589 &  11.119  &  1.739  &   .997  &  3.283\cr  
       .150 &  -2.352 &   3.577 &  11.681  &  1.820  &  1.033  &  3.381\cr 
       .120 &  -2.620 &   3.554 &  12.512  &  1.975  &  1.079  &  3.548\cr 
       .100 &  -2.979 &   3.510 &  13.685  &  2.284  &  1.138  &  3.751\cr  
       .098 &  -3.048 &   3.499 &  13.923  &  2.365  &  1.154  &  3.780\cr  
       .097 &  -3.087 &   3.493 &  14.056  &  2.409  &  1.162  &  3.791\cr 
       .096 &  -3.121 &   3.487 &  14.203  &  2.465  &  1.177  &  3.859\cr}}  

\par
The evolution of theoretical predictions concerning the HR diagram
location of VLM stellar models and the mass-luminosity relation
has been already discussed  by Baraffe et al. (1995, 
hereinafter BCAH95), Baraffe \& Chabrier (1996, hereinafter BC96) and Chabrier 
et al. (1996, hereinafter CBP96). BCAH95 presented a comparison 
between stellar models
based either on the $T(\tau)$ or on the {\sl "old"} generation 
of model atmospheres by  Allard \& Hauschildt (1995).
CBP96 discussed the effect of different treatments of the atmosphere on the
mass luminosity relation, presenting evolutionary computations which for 
the lower metallicities  still rely on the AH95 model atmospheres.
When revising this paper, improved models for metal-poor low-mass stars
have been presented by Baraffe et al. (1997, hereinafter BCAH97)
and for solar metallicity by Chabrier \& Baraffe (1997, hereinafter CB97),
so we have performed some additional comparisons with these updated models.

Figure 5 compares the HR diagram location of solar metallicity VLM models 
computed by relying either on B95 or on the Eddington approximation
with models presented by BCAH95,CBP96 and CB97. One finds that present
results distribute in between the "old" (BCAH95) and the "new" generation
models (CBP96) presented by the group of Baraffe and coworkers. This is
a rather surprising result, since present models and  CB97 should have quite
a similar input physics (following the results discussed by BC96),
and it is not clear to us where the difference
is coming from. Nor the origin of the (small) differences between
BC96 and CB97 has been till now discussed in the literature. 
One also finds the curious evidence that most updated
CB97 models appear in close agreement with our models computed 
with  $T(\tau)$ boundary condition.
\figure{5}{S}{80mm}{\bf Figure 5. \rm The location in the HR diagram of 
present stellar models for solar metallicity compared with 
similar models but from  BCAH95, BC96 and CB97. 
}

The main difference between the NG97 models (Allard \& Hauschildt 1997, 
Leggett et al. 1996, CBP96) and the B95 ones is the adoption
of different line lists for $\rm TiO$ and $\rm H_2O$ (by J\"orgensen 
1994 in NG97 and by Plez et al. 1992 in B95).
However, BC96 (and also CBP96) 
compared stellar models computed with their evolutionary code
using alternatively the B95 model atmospheres and the NG97 ones, obtaining only
negligible differences.
As discussed by Bessel (1995, 1996), one has to 
remind that for temperatures cooler
than $3000K$ B95 models have much too strong $\rm H_2O$ bands, indicating
that  the $\rm H_2O$ opacity is overestimated. Such occurrence should
have the effect of decreasing the B95 fits of
spectra of the coolest M dwarfs by about $100K$ (Bessel 1995),
the consequences on the HR diagram location of the models being 
correctly valuable only when Brett model atmospheres with 
updated  $\rm H_2O$ opacities will be available. However, 
the differences between present and CB97 models largely occur
at effective temperatures larger than $3000K$, thus the discrepancies
can be hardly ascribed to such an effect. Here we can only conclude that
such differences deserve further investigations, data in 
Figure 5  giving an indication of the degree of freedom still
existing in theoretical predictions.

\par
As far as the mass-luminosity relation is concerned, CBP96 have already
discussed the effect of non-gray model atmospheres on the reliability
of {\sl m-L} relation, and that discussion will not be
repeated here. Moreover, Kroupa \& Tout (1997) (but see also von Hippel et al. 1997)
have recently presented - still
relying on $T(\tau)$ stellar models - an investigation on the metallicity
dependence of the theoretical mass - magnitude relation.
Figure 6 shows the {\sl m-$M_V$} and {\sl m-$M_I$} relations
for solar metallicity VLM objects, as obtained by using $T(\tau)$ or 
B95 atmospheres. For the sake of comparison we report in the same
figure similar predictions from BC96, with magnitudes derived from the
published luminosities according to the  above quoted 
procedure. Figure 7 shows the mass-luminosity relations 
in selected  photometric bands for the {\sl "best"} models 
at the two lower metallicities investigated in the present work. As a
relevant point, one finds that the {\sl m-$M_K$} relation appears 
scarcely  affected by metallicity effect, at least for $M > (0.11 - 
0.12)M_{\odot}$, i.e. for $M_K < 9.0$mag.
Such occurrence could be of some help when planning 
observational surveys devoted to investigated the mass function for field
stars, whose metallicity is usually not well known.

\par
In panel c) of the previous Figure, we report also the semi-empirical {\sl m-$M_I$} 
relation given  by Fahlman et al. (1989) for a metallicity $Z\approx0.0001$.
One finds that such an estimate appears in rather good agreement with 
present theoretical predictions for {\sl m-$M_I$} at Z=0.0002 
for $M>0.16M_{\odot}$ i.e. $M_I<10$mag. At larger magnitudes, 
the semi-empirical relation crosses our predictions for Z=0.0002,
predicting a larger magnitude for a given stellar mass. 
However, at this faint end of the
{\sl m-$M_I$} relation, Fahlman et al. (1989) adopted theoretical 
models by D'Antona (1987), computed by using "old" physics both 
for the EOS and the low temperature
opacity, possibly affecting the semi-empirical results for $M_I$ 
magnitudes above $\approx(9-10)$mag.

\par
Since  the  first derivative of the mass-magnitude relation
is a key tool in interpreting observed luminosity functions
in term of the mass function, previous results for  the various bands
have been best fitted to obtain simple  analytical relations of the
form:

$$\log{M/M_{\odot}}= a_0 + a_1\cdot{M_x} + a_2\cdot{M^2_x} + a_3\cdot{M^3_x} + a_4\cdot{M^4_x}$$
\noindent
Table 4 shows the values of the coefficients for the different 
metallicities together with the value of the standard deviation
$\sigma$ for each relation.
\table{4}{D}{\bf Table 4. \rm Coefficients of the polynomial regression for the {\sl mass -
magnitude} relation, for the various metallicities and photometric bands adopted
in the present work. The last column lists the standard deviation.} 
{\halign{%
\hfil\rm# & \hskip10pt\rm#\hfil  & \hskip10pt\rm#\hfil &\hskip10pt\hfil\rm#
&\hskip10pt\hfil\rm# & \hskip10pt\hfil\rm\hfil# &\hskip10pt\hfil\rm\hfil#\hfil&\hskip10pt\hfil\rm\hfil# & \hskip5pt\hfil\rm\hfil# \cr
$Z$ & $M_x$ &  & $a_0$  & $a_1$ & $a_2$ & $a_3$ & $a_4$ & $\sigma$ \cr 
\noalign{\vskip 10pt}
0.02 & $M_V$ & $<12.5\rm mag$ & 39.76505 & -15.11174  & 2.13080  & -0.13239  &  3.03803E-3 & 0.003 \cr
     & $M_V$ & $>12.5\rm mag$ & 6.22027  &  -1.09083  & 0.05592  & -9.81080E-4  &             & 0.001 \cr
     & $M_I$ & $<10\rm mag$ &  28.51610  & -13.85940  &  2.50028  & -0.19905  & 5.84703E-3 & 0.002 \cr
     & $M_I$ & $>10\rm mag$ & -14.72470  &   5.59494  & -0.77702  &  4.56054E-2  &  -9.75915E-4 & 0.0005 \cr
     & $M_R$ & $<11\rm mag$ &  33.60587  & -14.47284  & 2.31192  & -0.16278   &  4.23001E-3   & 0.003 \cr
     & $M_R$ & $>11\rm mag$ & -2.94907  &  1.45823   & -0.22239  & 1.27567E-2   &  -2.55325E-4   & 0.0008 \cr
     & $M_K$ & $<7.5\rm mag$ &  18.32809 & -11.93366  & 2.88362  & -0.30845   &  1.21653E-2 & 0.002 \cr
     & $M_K$ & $>7.5\rm mag$ & -2.70027 & 1.17979 & -0.17640 & 7.51283E-3  &  & 0.0002 \cr
\noalign{\vskip 5pt}
0.002& $M_V$ & $<14\rm mag$ &  4.40273 & -2.11497 & 0.36603 & -2.77915E-2 & 7.49083E-4  & 0.013 \cr
     & $M_V$ & $>14\rm mag$ & 65.05768  & -16.25213   & 1.50485 & -6.21070E-2 & 9.62896E-4  & 0.0002 \cr
     & $M_I$ & $<9.0\rm mag$ & 8.10679  & -4.42546  &  0.87922 &  -7.64219E-2   & 2.37850E-3  & 0.004 \cr
     & $M_I$ & $>9.0\rm mag$ & -19.06139 & 7.46401 & -1.07565 & 6.59578E-2 & -1.47510E-3 & 0.0009 \cr
     & $M_R$ & $<9.0\rm mag$ & -0.49015  & 0.67463  & -0.22982 & 2.84768E-2 & -1.24476E-3  & 0.0004 \cr
     & $M_R$ & $>9.0\rm mag$ & -0.54928  & 0.67247  & -0.14004 & 9.20567E-3 & -2.00095E-4 & 0.002 \cr
     & $M_K$ & $<7.0\rm mag$ & 7.56689  & -5.02387  & 1.22199 & -0.13146  & 5.11053E-3  & 0.001 \cr
     & $M_K$ & $>7.0\rm mag$ & 0.37826   & -0.05447 & 1.69223E-3 & -3.70568E-3  & 2.68112E-4  & 0.001 \cr
\noalign{\vskip 5pt}
0.0002 & $M_V$ & $<10\rm mag$   & 3.81682  & -1.71337 & 0.26601 & -1.71522E-2 & 3.31352E-4 & 0.002 \cr
       & $M_V$ & $>10\rm mag$   & 0.86197  &  3.29285E-2 & -3.16077E-2 & 1.40673E-3 &             & 0.002 \cr
       & $M_I$ & $<8.5\rm mag$  & 1.97179 & -0.96217 & 0.15045 & -8.53940E-3 &   & 0.005 \cr
       & $M_I$ & $>8.5\rm mag$  & -1.36246 &  0.68103 & -0.10568 & 4.27400E-3 &             & 0.001 \cr
       & $M_R$ & $<8.5\rm mag$ &  -9.55902 &  5.94358 & -1.37655 & 0.13923 & -5.25802E-3 & 0.001 \cr
       & $M_R$ & $>8.5\rm mag$ &   2.08701 & -0.31581 & -4.86486E-3 & 8.30517E-4 &             & 0.002 \cr
       & $M_K$ & $<7.0\rm mag$ &   1.21849 & -0.68787 & 0.12464 & -9.01619E-3 &             & 0.005 \cr
       & $M_K$ & $>7.0\rm mag$ & -10.15714 &  5.01276 & -0.91221   & 6.91446E-2 & -1.88977E-3 & 0.002 \cr}}
\figure{6}{S}{80mm}{\bf Figure 6. \rm {\sl m - $M_V$} and {\sl m - $M_I$} relations
for solar metallicity, 10Gyr old models, for various assumptions concerning the outer boundary
conditions (see text).}
\figure{7}{S}{100mm}{\bf Figure 7. \rm As in Figure 6 but for Z=0.002 
and Z=0.0002. The mass-luminosity relation is shown only for {\sl "best"} 
models in selected photometric bands. Panel c) shows also the {\sl m-$M_I$} relation
from Fahlman et al. (1989).}
\section{Theory {\sl versus} observations.}
\tx

In Paper I we found that a rather satisfactory agreement between observational
data and theoretical VLM models can be  achieved even by relying
on models based on the Eddington approximation. 
Let us here compare present "best" models with observations.

\figure{8}{S}{80mm}{\bf Figure 8. \rm (V, V-I) CM diagram for the lower main sequence
of NGC6397 (Cool et al. 1996) as compared with theoretical isochrones 
for [M/H]=-2.04 , and for the ages 10, 12 and 14Gyr (Cassisi et al. 1997) shifted
to account for a cluster distance modulus and reddening $(m-M)_V=12.50$ 
and $E(V-I)=0.20$. The MS locus for VLM structures, for [M/H]=-1.04 t=10 Gyr is also shown.} 
\par
The most relevant observational sample  is obviously 
represented by recent  Hubble Space Telescope data 
for lower main sequences in galactic globular clusters.
Several CM diagrams  have been already presented (NGC6397: Paresce, 
De Marchi \&
Romaniello 1995, Cool, Piotto \& King 1996 and, Mould et al. 1996; 
47Tuc, M30: King, Cool \& Piotto 1996, Piotto, Cool \& King 1997;
 NGC6752: Ferraro et al. 1997; NGC6656: De Marchi \& Paresce 1997).
The most tight sequence of VLM stars
in a GC appears the one presented by Cool et al. (1996), which represents 
a fundamental tool to test  the theory of VLM structures.
In Paper I  it has been shown that a largely
satisfactory agreement has been achieved between our $T(\tau)$ 
models and observation.
In figure 8, we perform the same comparison but using present {\sl best} VLM
models for metallicity [M/H]=-2.04 and -1.04, adopting from Alcaino et 
al. (1987) a cluster distance modulus $(m-M)_V=12.50$ and a 
reddening $E(V-I)=0.20$. Data in figure 8
have been implemented at larger luminosities with isochrones for the ages
10, 12 and 14Gyr (Cassisi et al. 1997), as computed by adopting 
the same opacity
evaluations used in this work but the OPAL equation of state 
(Rogers, Swenson \& Iglesias 1996)
to allow the required match with the VLM sequence (see Brocato, 
Cassisi \& Castellani 1997a for a discussion on that matter).

\par
One finds that observational data
agree fairly well with present theoretical predictions for  metal
poor models. Interesting enough, one can notice that not reasonable
fitting can be achieved with the moderately metal rich sequence 
shown in the same figure. Thus theoretical results appear in good agreement
with  current estimates for the cluster metallicity, namely 
[M/H]$\simeq-1.61$, where an enhancement of $\alpha$ elements 
by $[\alpha/Fe]=0.30$ has been taken into account (see Brocato, Cassisi
\& Castellani 1997b for more details). As shown in the same Figure, some 
residual discrepancies between theory and
observations still exist, to be eventually better understood  but only 
when updated theoretical color - $T_{eff}$ relations suitable for 
metal poor stars will become available (Allard et al. 1997).

\par
Figure 9 shows the most complete presently available CM diagram 
for stars with known parallaxes, as obtained implementing the sample provided by
Monet (1992) with recent data by Dahn et al. (1995).
In Paper I, it has been already shown that
metal poor $T(\tau)$ models  and  BCAH95 models for [M/H]=-1.5 
appear in rather good agreement with the location of the hotter subdwarfs 
sequence. However, that paper also disclosed that all theoretical 
models, including BCAH95, failed in accurately
reproducing the location of fainter objects for the cooler sequence of stars,
usually interpreted as the sequence of VLM stars with solar metallicity.
To test if the use of more accurate outer boundary 
conditions can help in reducing the
discrepancy between observations and theory,  
the same Figure 9 gives the location
of {\sl "best"} models, for the three metallicities adopted in this work,
together with the $T(\tau)$ sequence of models for solar 
metallicity from Paper I and the
BCAH95 and BC96 models (all for solar metallicity).

\figure{9}{S}{80mm}{\bf Figure 9. \rm $M_V$ versus $(V-I)$ diagram for faint stars
with known parallaxes as provided by Monet et al (1992) or Dahn et al. (1995).
Theoretical predictions from this paper for the  [M/H]=0.0, -1.04
and -2.04, are also displayed together with the theoretical models by BCAH95, BC96
and the $T(\tau)$ models for solar metallicity from Paper I.}
An inspection of the figure leads to the following conclusions:
\medskip
\noindent
i) {\sl "best"} metal poor models rank very well along 
the hotter subdwarfs sequence, supporting the indication given in Paper I
about the lower limit for the metallicity of disk subdwarfs and 
the evidence that the CM diagram location
of VLM stars appears as {\sl a metallicity indicator of unusual sensitivity} 
(Paper I, Brocato, Cassisi \& Castellani 1997a, 1997b);
\smallskip
\noindent
ii) B95 models significantly improve the fit of the metal rich sequence 
in comparison
with {\sl old} $T(\tau)$ models. Present models appear also in best 
agreement with observation with
respect to BC96 models in the color range $2.2\le(V-I)\le3.0$, where
the main sequence location is strongly affected by the adopted 
treatment for the outer boundary
conditions (as discussed in Paper I and in BCAH95).
Nevertheless, one finds that a significant discrepancy still exists
for colors $(V-I)>3.0$ mag, i.e. $M_V\ge14$ mag;
\smallskip
\noindent
iii) Curiously enough, BCAH95 models for solar metallicity 
seem to match the location
of the cooler sequence in the CM diagram better than 
both present {\sl "best"} and BC96 models. 
Since both present and  BC96 models have been computed by adopting
a treatment of the atmosphere much more accurate than in 
BCAH95, this occurrence
has to be perhaps regarded as an evidence that some 
other {\sl "ingredient"} (opacity?, color-temperature relation?),
adopted in the computations, needs further improvements.
\medskip
\noindent
\figure{10}{S}{80mm}{\bf Figure 10. \rm The HR diagram location 
of VLM object from Leggett et al. (1996) as compared with theoretical 
predictions from the present paper or BCAH97 for the various metallicities,
as labeled. Solar metallicity models from BCAH95 and BC96 are also
displayed.}
Let us finally compare our VLM stellar models with observational data
recently provided by Legget et al. (1996), who investigated 
16 red dwarfs providing bolometric luminosities and  
effective temperatures. These authors already found a good agreement
between their data and BCAH95 models. Figure 10 now
compares Legget et al. (1996) data with present {\sl "best"} and BCAH97
predictions for various assumptions about the metallicity, and with
BCAH95, BC96 and CB97 solar metallicity models. Inspection of the Figure shows
that within the accuracy of observational data one may only 
conclude for a general agreement between theories and observation, 
without making a choice among the different approaches. 
In this context, it is worth noting that the 
object Gl 65AB which tends to be cooler than the solar metallicity
sequence is  an unresolved binary
system. If this system would resolved into two similar components, 
these would lie near the point marked by  an arrow (Leggett et al. 1996), 
in fine agreement with the theoretical prediction.
\figure{11}{S}{80mm}{\bf Figure 11. \rm Mass-luminosity relations for the labeled
assumptions on the stellar metallicity, as obtained in this work, by BCAH95 and BC96.
Observational data have been provided by Henry \& McCarthy (1993).}

\par
Figure 11 compares  theoretical {\sl m-$M_V$} 
relations from the present paper and from BCAH95 and BC96—
with observational data for nearby binaries by 
Henry \& McCarthy (1993). The two objects at $M_V\simeq8.6$ mag (listed as
Gl 677A and Gl 677B) which deviate significantly 
from theoretical prescriptions, belong to the same
binary system, with no large accuracy of the orbital parameters   since 
the orbit has been followed for less than a revolution (Henry \& McCarthy 1993)
Again one finds that theory and observations
appear in satisfactory agreement, without allowing a choice among the 
different theoretical approaches.

\par
Figure 12 finally compares the mass luminosity function given by
Kroupa et al (1993) with present and previous computations on the
matter (for solar metallicity), as labeled. Now one finds that the most recent results appear
in better agreement, without any clear indication allowing a choice
between present models and CB97. In the above quoted paper (and reference 
therein) Kroupa et al. discussed also the relevance of the run with  $M_V$ of the 
derivative of the mass-luminosity function $dm/dM_V$.
Therefore, we have decided to compare the derivative of the mass - luminosity
function obtained by adopting the most updated VLM stellar models, presently
available, with the derivative of Kroupa et al.'s (1993) relation, as given
by Kroupa \& Tout (1997). Figure 13
shows that present models foresee a maximum in the derivative whose 
location appears in better agreement with observational prescriptions
than CB97 models do. Thus in this respect it appears that present models 
work better.
\figure{12}{S}{70mm}{\bf Figure 12. \rm Mass - absolute visual magnitude relations
for solar metallicity stellar models. Present results are plotted as solid line, models by
CB97 as a dashed curve. Models of BCAH95 and BC96 are plotted as open squares
and open triangles, respectively. Observational data correspond to the empirical
$m - M_V$ relation derived by Kroupa, Tout \& Gilmore (1993).}
\figure{13}{S}{65mm}{\bf Figure 13. \rm The derivative of the $m - M_V$ relation as function
of $M_V$. The solid curve is the derivative of present models and the dashed line corresponds
to the derivative obtained adopting the CB97 stellar models. The observational data
correspond to the derivative of the mass - luminosity relation provided by Kroupa
\& Tout (1997).}

\section{Conclusions.}
\tx
In Paper I, we investigated the effects of new physical 
inputs as equation of state
and low temperature opacities on the evolutionary properties 
of VLM stellar objects,
by computing stellar models relying on the Eddington approximation for 
the treatment of outer boundary conditions.
In this work, we  devoted our attention to the effect of improving
the treatment of the atmosphere, by adopting  Brett's (1995a,b) non-grey model
atmospheres. 

\par
Present models have been compared with other VLM stellar 
models  in the current literature, as given by BCAH95, BC96, BCAH97
and CB97.
Comparison  with observational data for metal poor GC stars or
solar neighborhood dwarfs shows a reasonable agreement. 
This result can be regarded as a plain 
evidence that the solution of the long-standing discrepancy between theoretical
predictions and observed HR diagram location of VLM stars is no
more a tantalizing goal. However, a not negligible
discrepancy still exist between stellar models and CM location of the cooler
sequence of VLM objects with known parallaxes, for magnitudes larger than $M_V>14$mag.
This occurrence and the presence of still significant uncertainties 
in the observational
data - in particular concerning the effective temperature - confirms (as pointed
out by Legget et al. 1996, Bessell 1995, CBP96) that there
are still improvements to be performed, both in the models (structural and atmospheric)
and in the observations.

\section*{Acknowledgments}

\tx 
We gratefully acknowledge M.S. Bessell for providing us with useful
informations and suggestions about updated non gray model atmospheres.
J.M. Brett is also warmly thanked for providing with his updated model atmospheres.
We are also deeply in debt with C. van't Veer and F. Castelli for sharing with
us useful suggestions about the appropriate use of the Kurucz's model atmospheres.
G. Piotto, I.R. King and A.M. Cool are gratefully thanked for allowing us to use
their HST data. The anonymous referee suggested us to include the comparison with
the mass - luminosity relation by Kroupa et al. S.C. warmly thanks
G. Piotto for interesting discussions about luminosity and mass
function in globular clusters,
and S. Degl'Innocenti for the help provided at MPA.

\section*{References}

\bibitem Alcaino G., Buonanno R., Caloi V., Castellani V., Corsi C.E.,
 Iannicola G. \& Liller W. 1987, AJ 94, 917.

\bibitem Alexander D.R., Brocato E., Cassisi S., Castellani V., Degl'Innocenti S. \&
Ciacio F. 1997, A\&A 317, 90     

\bibitem Alexander D. R., Johnson H. R., \& Rypma R. L. 1983, ApJ 272, 773

\bibitem Alexander D.R. \& Ferguson J.W. 1994, ApJ 437, 879 

\bibitem Allard F. 1990, Ph.D. thesis, Univ. Heidelberg    

\bibitem Allard F. 1996, {\sl private comunication}

\bibitem Allard F., Alexander D.R., Hauschildt P.H. \& Schweitzer A. 
     1997, ApJ {\sl submitted to} 
     
\bibitem Allard F. \& Hauschildt P.H. 1994, in "Newsletter on Analysis of Astronomical
Spectra, ed. C.S. Jeffery (Daresbury Laboratory, UCL), p.9   

\bibitem Allard F. \& Hauschildt P.H. 1995a, ApJ 445, 433 

\bibitem Allard F. \& Hauschildt P.H. 1995b, in Proceedings of the ESO workshop   
 "The bottom of the Main Sequence  and Beyond", ed. Tinney C.G., p.32   

\bibitem Allard F. \& Hauschildt P.H. 1997, {\sl in preparation}    

\bibitem Auman J.R. Jr. 1969, ApJ 157, 799  

\bibitem Baraffe I. \& Chabrier G. 1995, in Proceedings of the ESO workshop   
 "The bottom of the Main Sequence  and Beyond", ed. Tinney C.G., p.24   

\bibitem Baraffe I. \& Chabrier G. 1996, ApJ 461, L51 

\bibitem Baraffe I., Chabrier G., Allard F. \& Hauschildt P.H. 1995, ApJ 446, L35

\bibitem Baraffe I., Chabrier G., Allard F. \& Hauschildt P.H. 1997,A\&A {\sl submitted to}

\bibitem Bessel M.S. 1995, in Proceedings of the ESO workshop    
 "The bottom of the Main Sequence  and Beyond", Tinney C.G., ed., p.123   
 
\bibitem Bessel M.S. 1996, {\sl private comunication}  

\bibitem Brett J.M. \& Plez B. 1993, Proc. ASA 10, 250  

\bibitem Brett J.M. 1995a, A\&A 295, 736     
 
\bibitem Brett J.M. 1995b, A\&ASS 109, 263      

\bibitem Brocato E., Cassisi S. \& Castellani V. 1997a, to appear in: 
"Advances in Stellar Evolution", ed. R.T. Rood \& A. Renzini (Cambridge Press)  

\bibitem Brocato E., Cassisi S. \& Castellani V. 1997b, {\sl in preparation}  

\bibitem Brocato E., Cassisi S., Castellani V., Cool A.M., King
I.R. \& Piotto G. 1996, in "Formation of the Galactic Halo... Inside
 and Out", ASP Conference Series, Vol. 92,
Morrison H. \& Sarajedini A., eds., p.76 

\bibitem Burrows A., Hubbard W.B. \& Lunine J.I. 1989, ApJ 345, 939 

\bibitem Burrows A., Hubbard W.B., Saumon D. \& Lunine J.I. 1993, ApJ 406, 158 

\bibitem Cassisi S., Castellani V., Degl'Innocenti S. \& Weiss A. 1997, A\&A {\sl submitted to}

\bibitem Chabrier G. \& Baraffe I. 1997, A\&A {\sl submitted to}

\bibitem Chabrier G., Baraffe I. \& Plez B. 1996, ApJ 459, L91 

\bibitem Cool A.M., Piotto G. \& King I.R. 1996, ApJ 468, 655  

\bibitem Cox A.N., Tabor J.E. 1976, ApJS 31, 271 

\bibitem D'Antona F. 1987, ApJ 320, 653 
  
\bibitem D'Antona F. \& Mazzitelli I. 1994, ApJS 90, 467  

\bibitem D'Antona F. \& Mazzitelli I. 1996, ApJ 456, 329   

\bibitem Dahn C.C., Liebert J., Harris H.C. \& Guetter H.H. 1995, 
in Proceedings of the ESO workshop "The bottom of the Main Sequence  and Beyond", Tinney C.G., ed., p.239   

\bibitem De Marchi G. \& Paresce F. 1997, ESO {\sl preprint} n. 1206

\bibitem Dorman B., Nelson L.A. \& Chau W. J. 1989, ApJ 342, 1003 
 
\bibitem Ezer D.W. \& Cameron A.G.W. 1967, Canadian J. Phys. 45, 3461  

\bibitem Fahlman G.G., Richer H.B., Searle L. \& Thompson I.B. 1989, ApJ 343, L49 

\bibitem Ferraro F.R., Carretta E., Bragaglia A., Renzini A. \& Ortolani S. 1997, MNRAS, 286, 1012

\bibitem Hayashi C. \& Nakano T. 1963, Prog. Theor. Phys. 30, 460 

\bibitem Henry T.O. \& McCarthy D.W. Jr, 1993, AJ 106, 773  

\bibitem J\o rgensen U.G. 1994, ApJ 284, 179  

\bibitem King I.R., Cool A.M. \& Piotto G. 1996, in "Formation of the Galactic
Halo... Inside and Out", ASP Conference Series, Vol. 92, Morrison H. \& Sarajedini A., eds., p.277  

\bibitem Krishna-Swamy K.S. 1966, ApJ 145, 174 

\bibitem Kroupa P., Tout C.A. \& Gilmore G. 1993, MNRAS 262, 545

\bibitem Kroupa P. \& Tout C.A. 1997, MNRAS {\sl in press}  

\bibitem Kui R. 1991, Ph.D. thesis, National University of Australia   

\bibitem Kurucz R.L. 1992, Rev. Mexicana. Astron. Astrof. 23, 223   

\bibitem Kurucz R.L. 1993, SAO CD-ROM   

\bibitem Leggett S.K., Allard F., Berriman G., Dahn C.C. \& Hauschildt P.H. 1996,
 ApJSS 104, 117     

\bibitem Limber D.N. 1958, ApJ 127, 387 

\bibitem M\'era D., Chabrier G. \& Baraffe I. 1996, ApJ 459, L87  

\bibitem Mihalas D. 1978, Stellar Atmosphere, 2d Ed. Freeman and Cie    

\bibitem Monet D.G., Dahn C.C., Vrba F.J., Harris H.C., Pier J.R.,
Luginbuhl C.B. \& Ables H.D. 1992, AJ 103, 638

\bibitem Morel P., van't Veer C., Provost J., Berthomieu G., Castelli F., Cayrel
R., Goupil M.J. \& Lebreton Y. 1994, A\&A 286, 91    

\bibitem Mould J.R. et al. 1996, PASP 108, 682

\bibitem Paresce F., De Marchi G. \& Romaniello M. 1995, ApJ 440, 216 

\bibitem Piotto G., Cool A.M. \& King I.R. 1997, AJ {\sl in press}

\bibitem Plez B., Brett J.M. \& Nordlund A. 1992, A\&A 256, 551   

\bibitem Rogers F.J., Swenson F.J. \& Iglesias C.A. 1996, ApJ 456, 902

\bibitem Salaris M. \& Cassisi S. 1996, A\&A 305, 858  

\bibitem Saumon D. 1994, in "The Equation of State in Astrophysics", IAU Symp
  n.147, Chabrier G. \& Schatzman E., eds., p.306 

\bibitem Saumon D., Bergeron P., Lumine L.I., Hubbard W.B. \& Burrows A. 1994, ApJ 424, 333   
     
\bibitem Saumon D. \& Chabrier G. 1992, Phys.Rev.A. 46, 2084 

\bibitem Saumon D., Chabrier G. \& Van Horn H.M. 1995, ApJS 99, 713 

\bibitem Tinney C.G. 1996, MNRAS 281, 644

\bibitem VandenBerg D.A., Hartwick F.D.A., Dawson P. \& Alexander D.R. 1983, ApJ 266, 747 

\bibitem von Hippel T., Gilmore G., Tanvir N., Robinson D. \& Jones D.H.P. 1996, AJ 112, 192

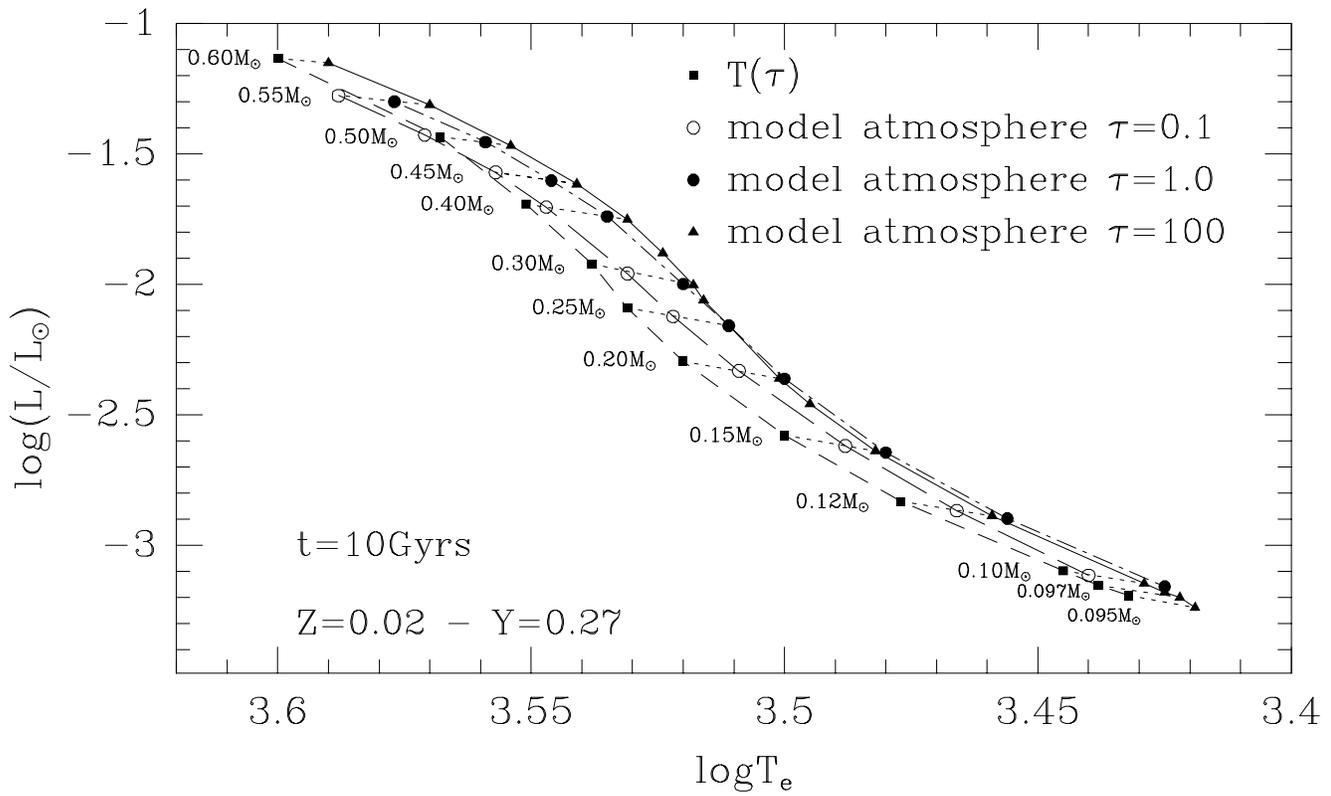
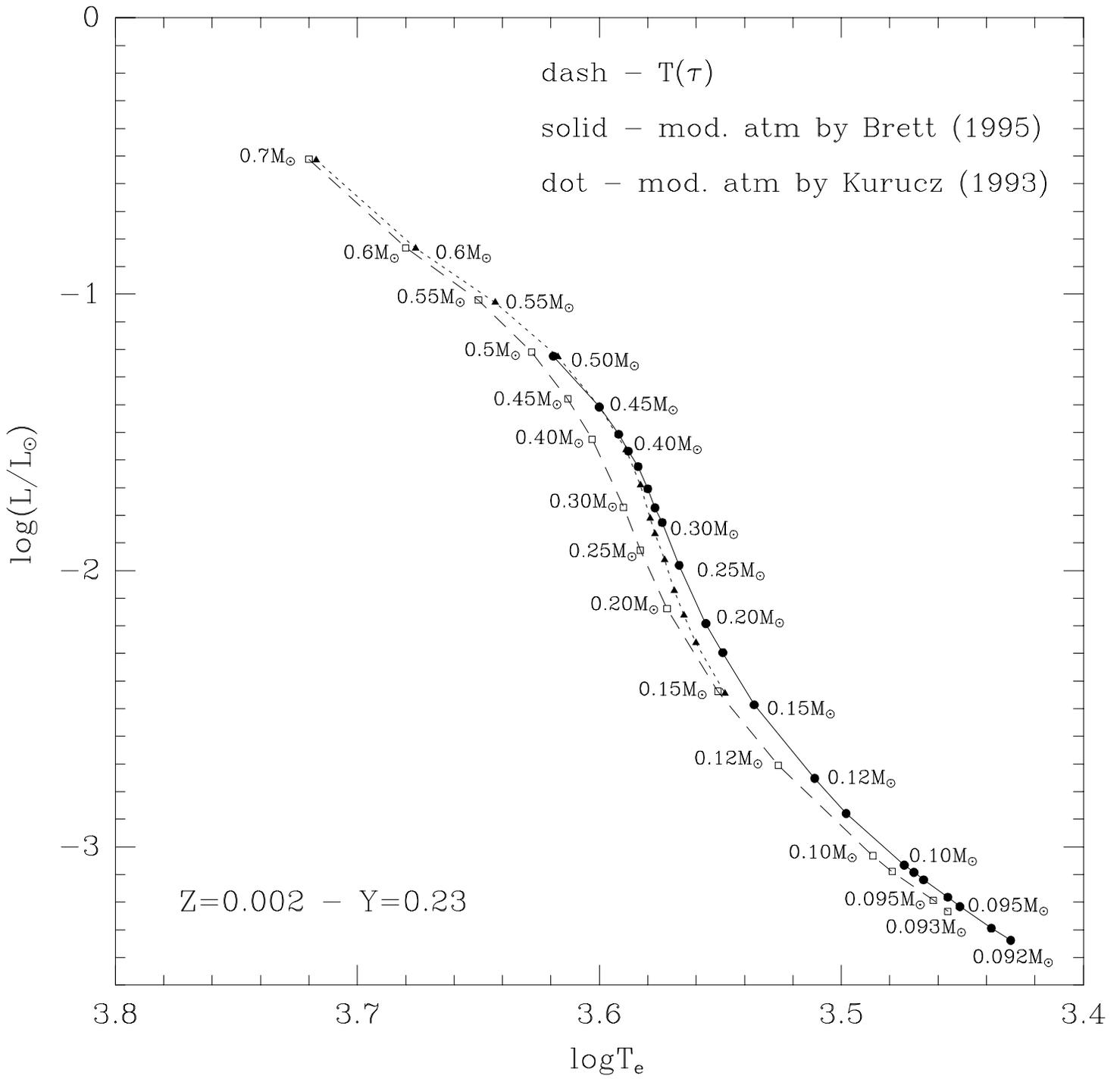
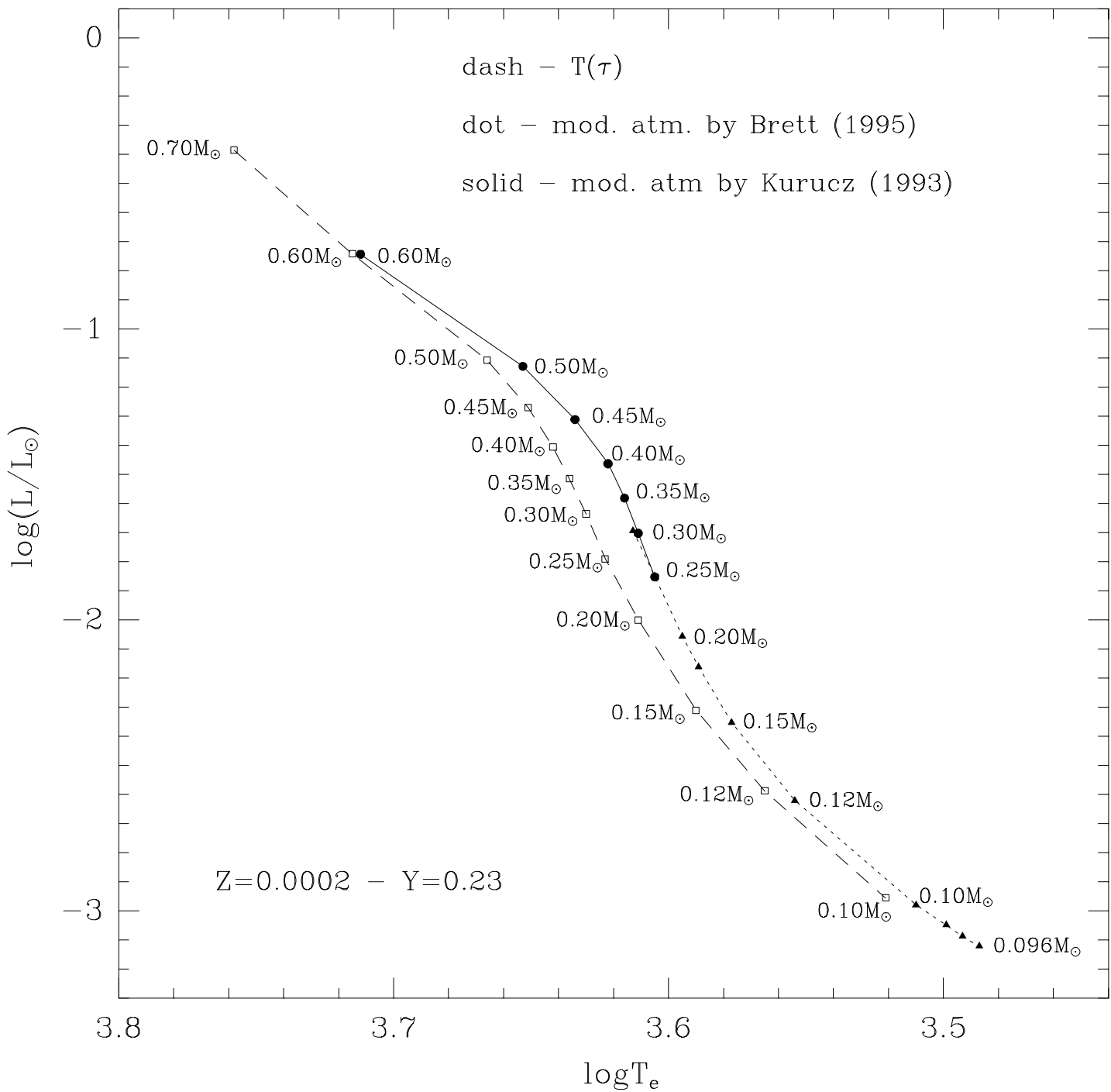
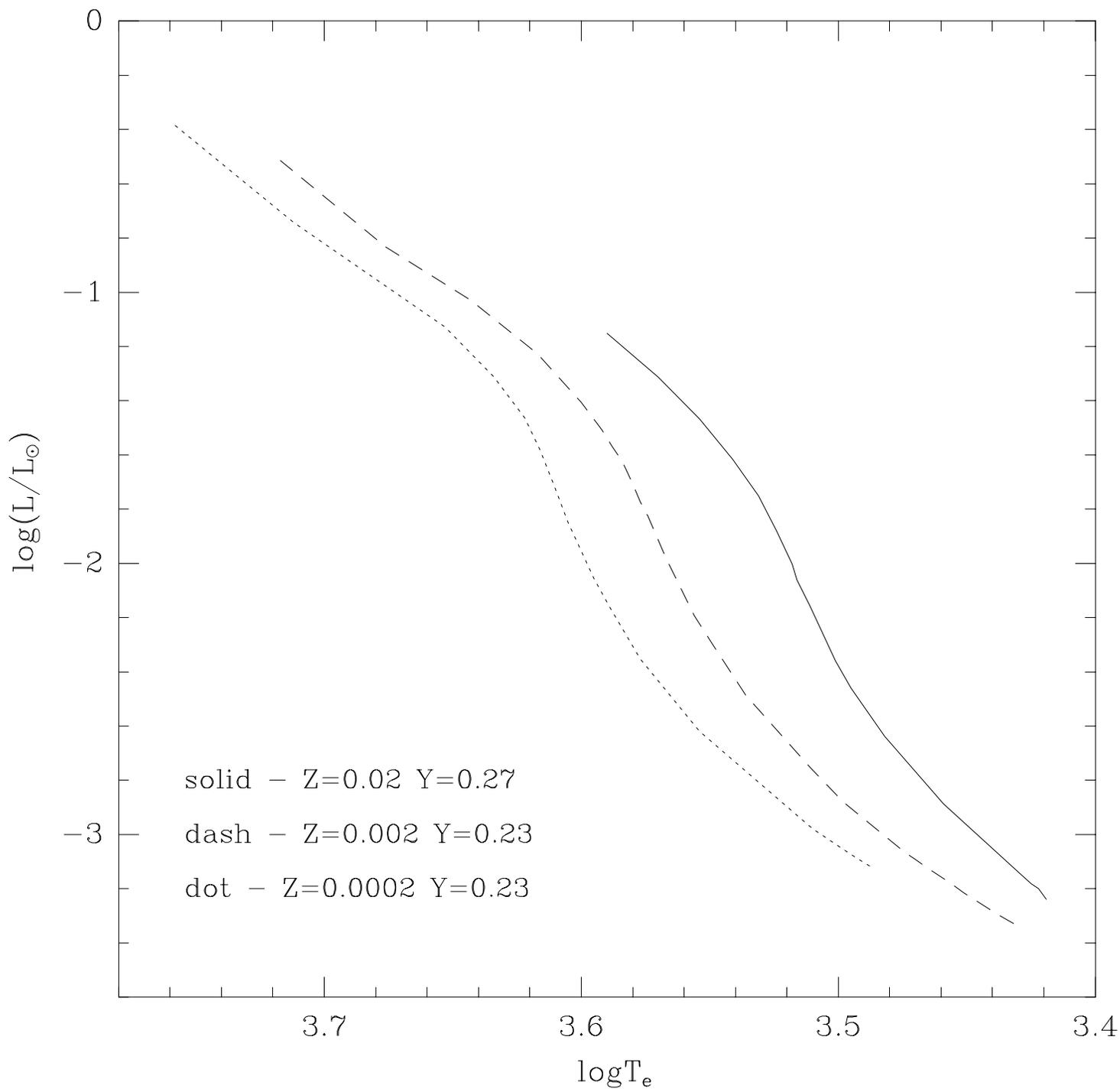
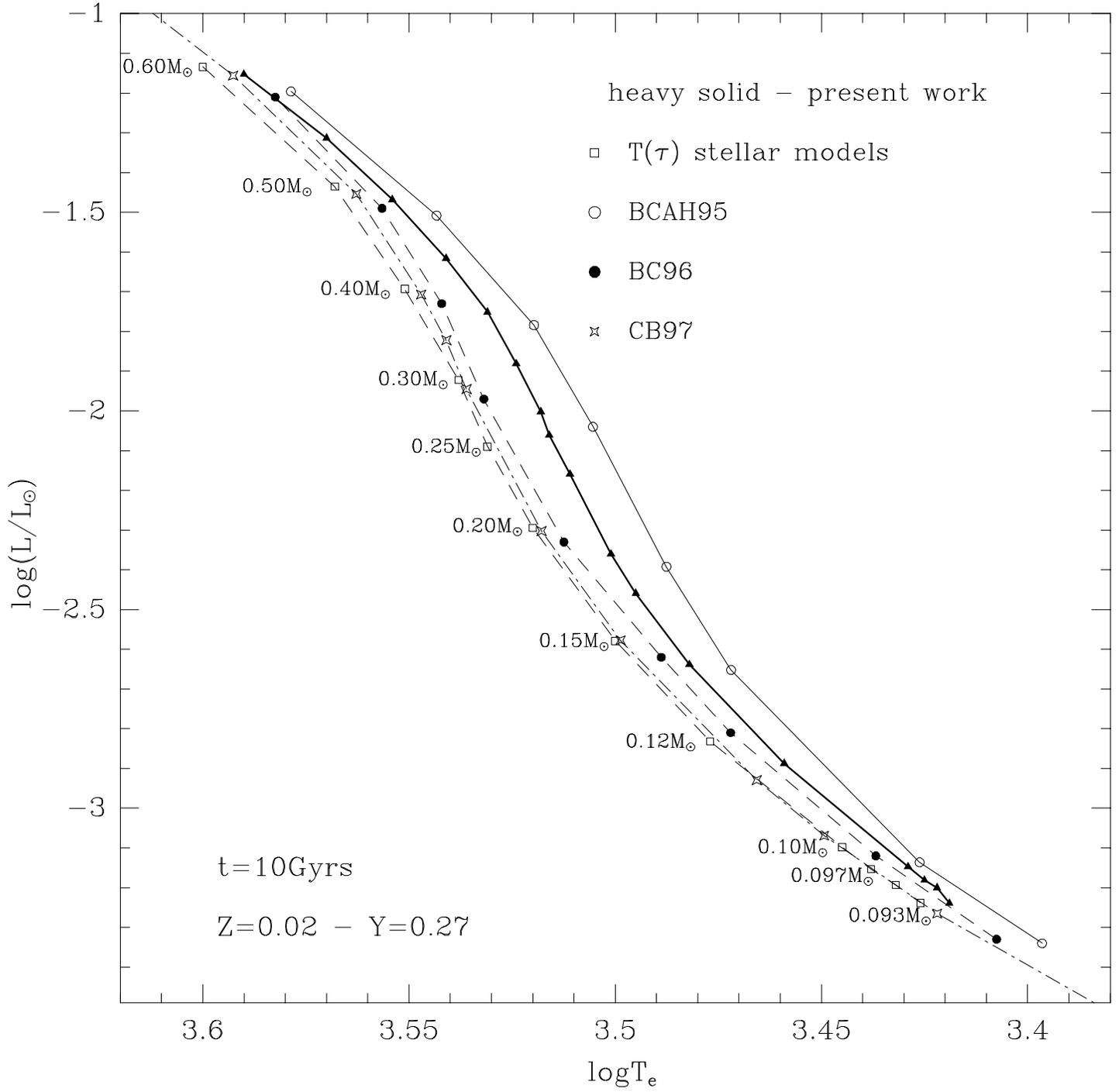
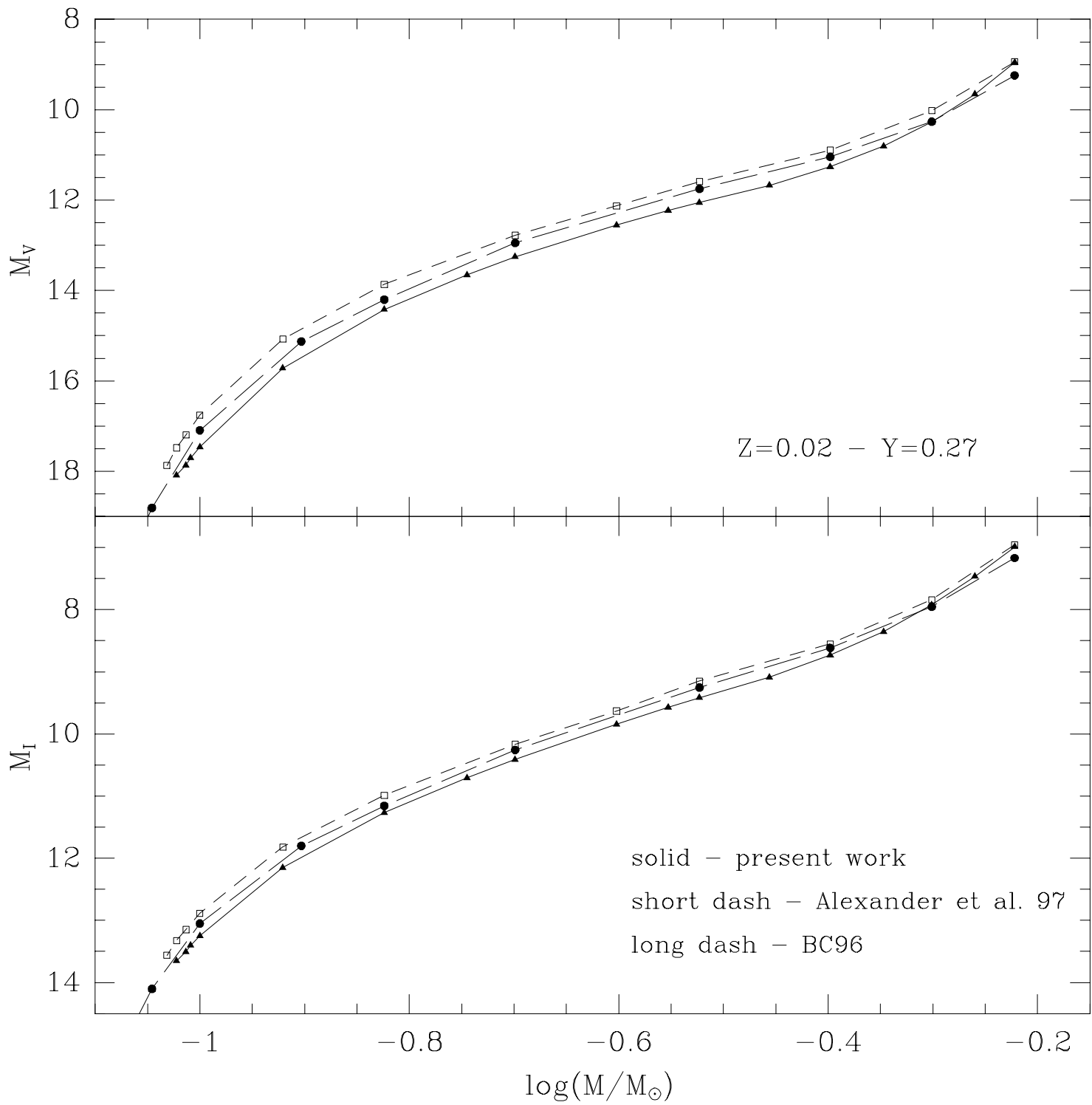
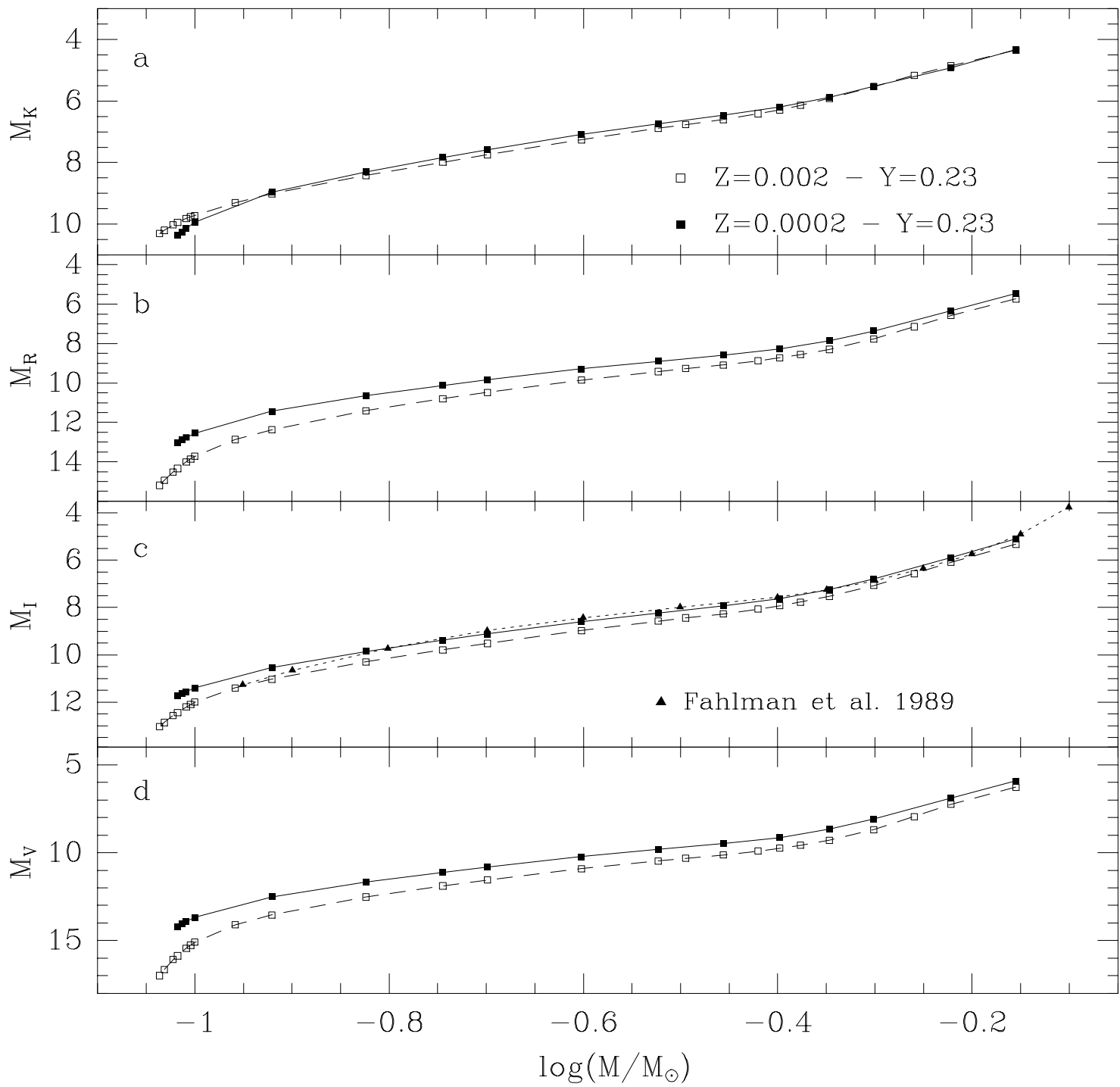
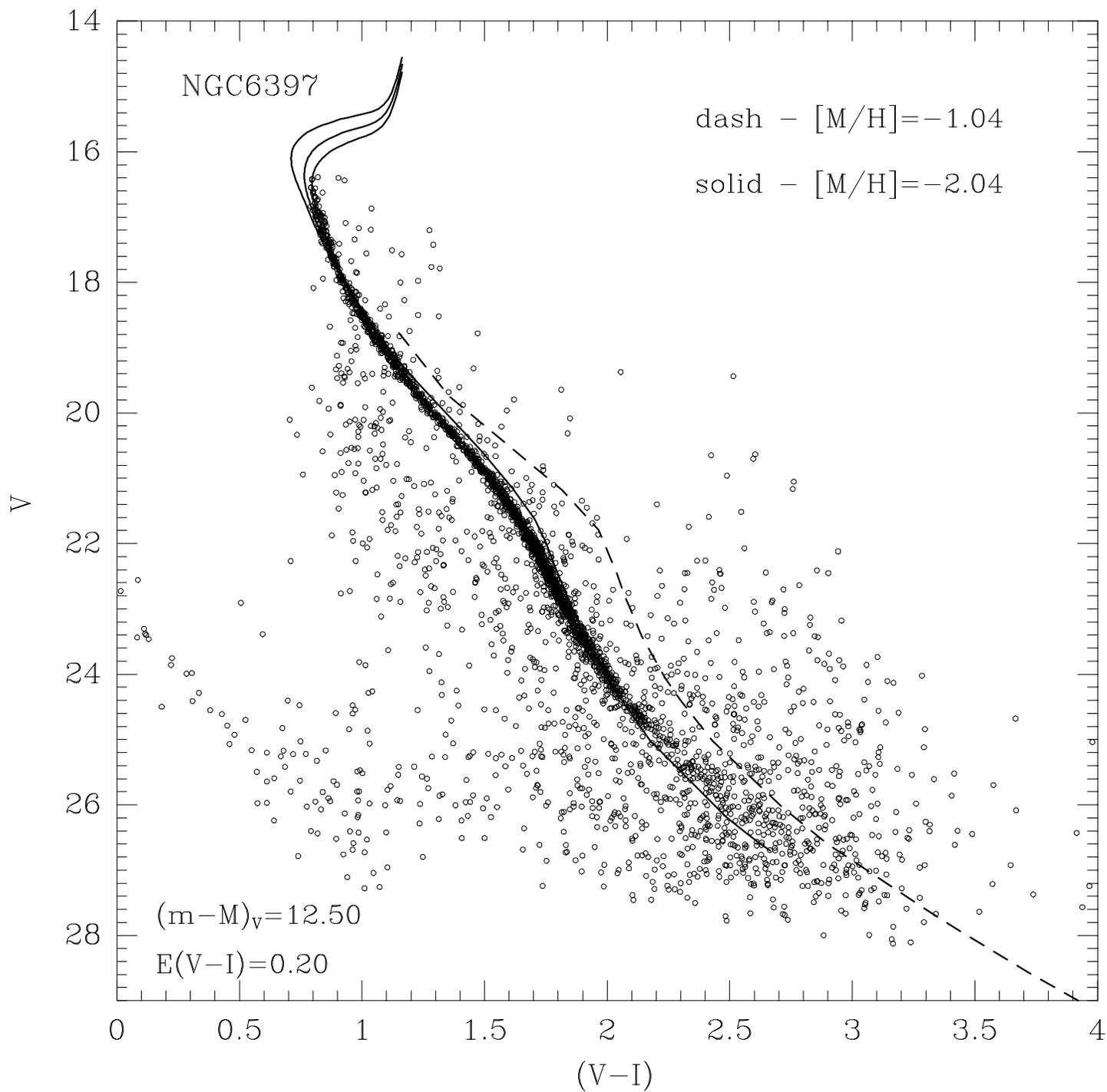
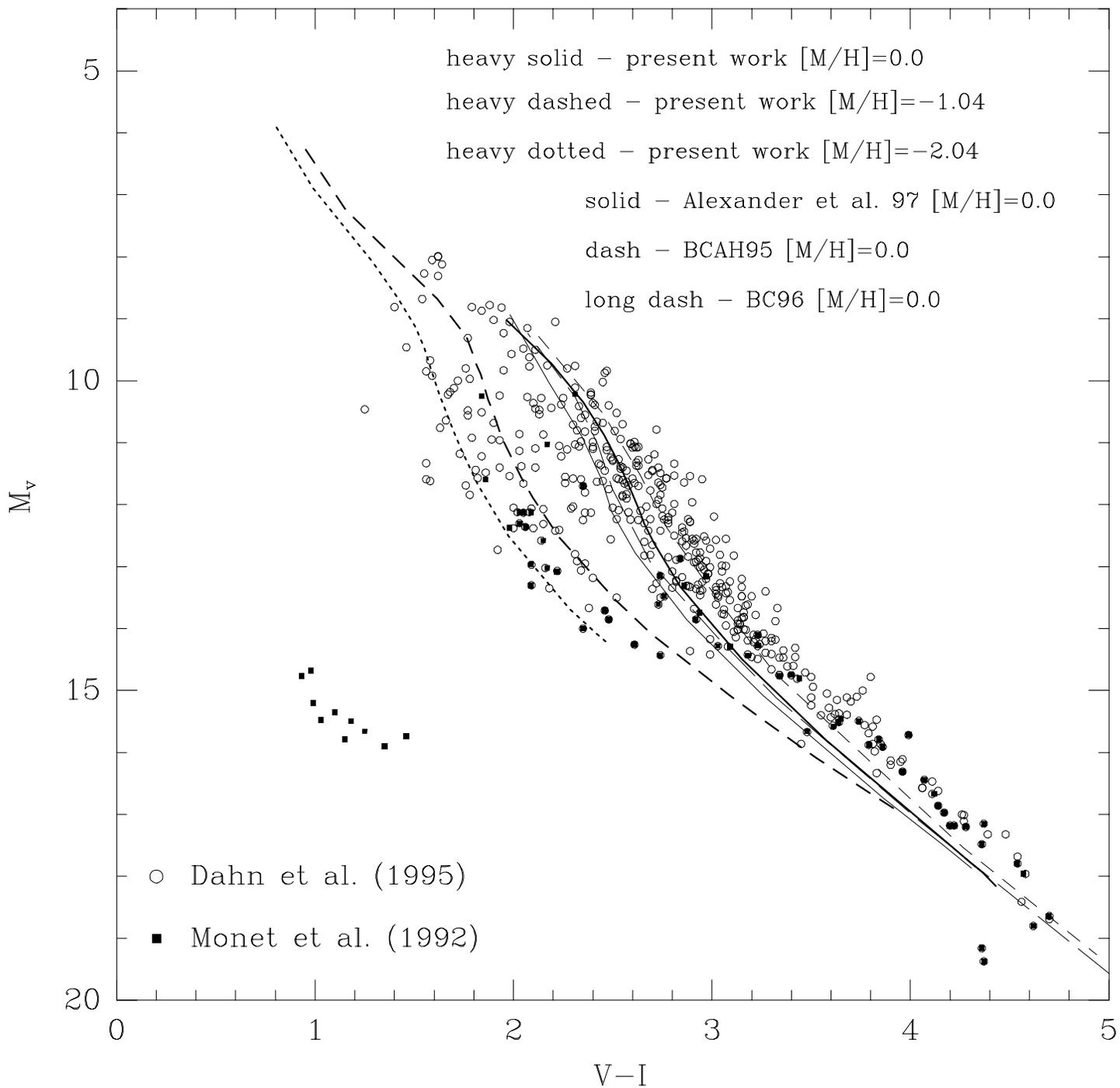
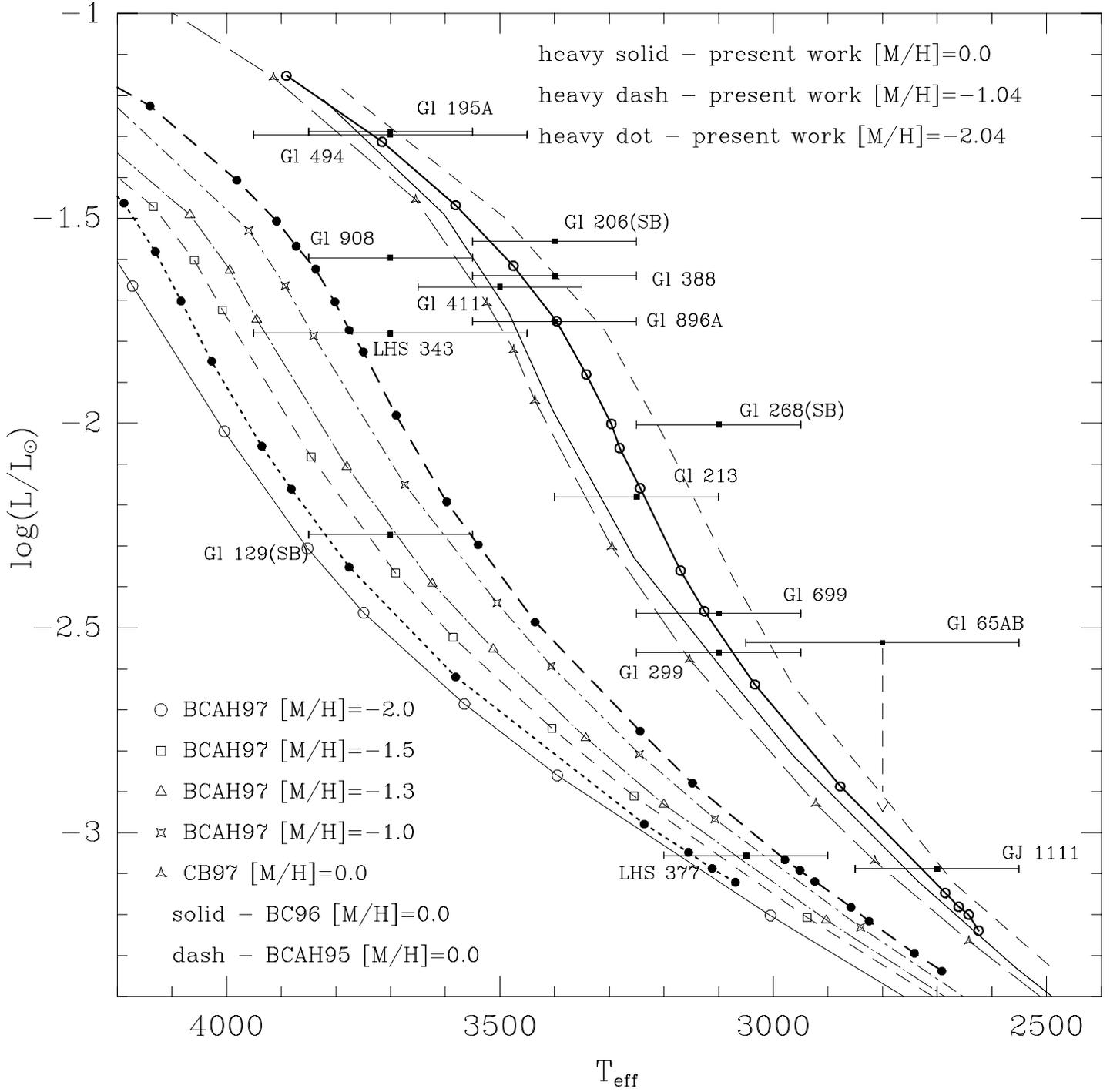
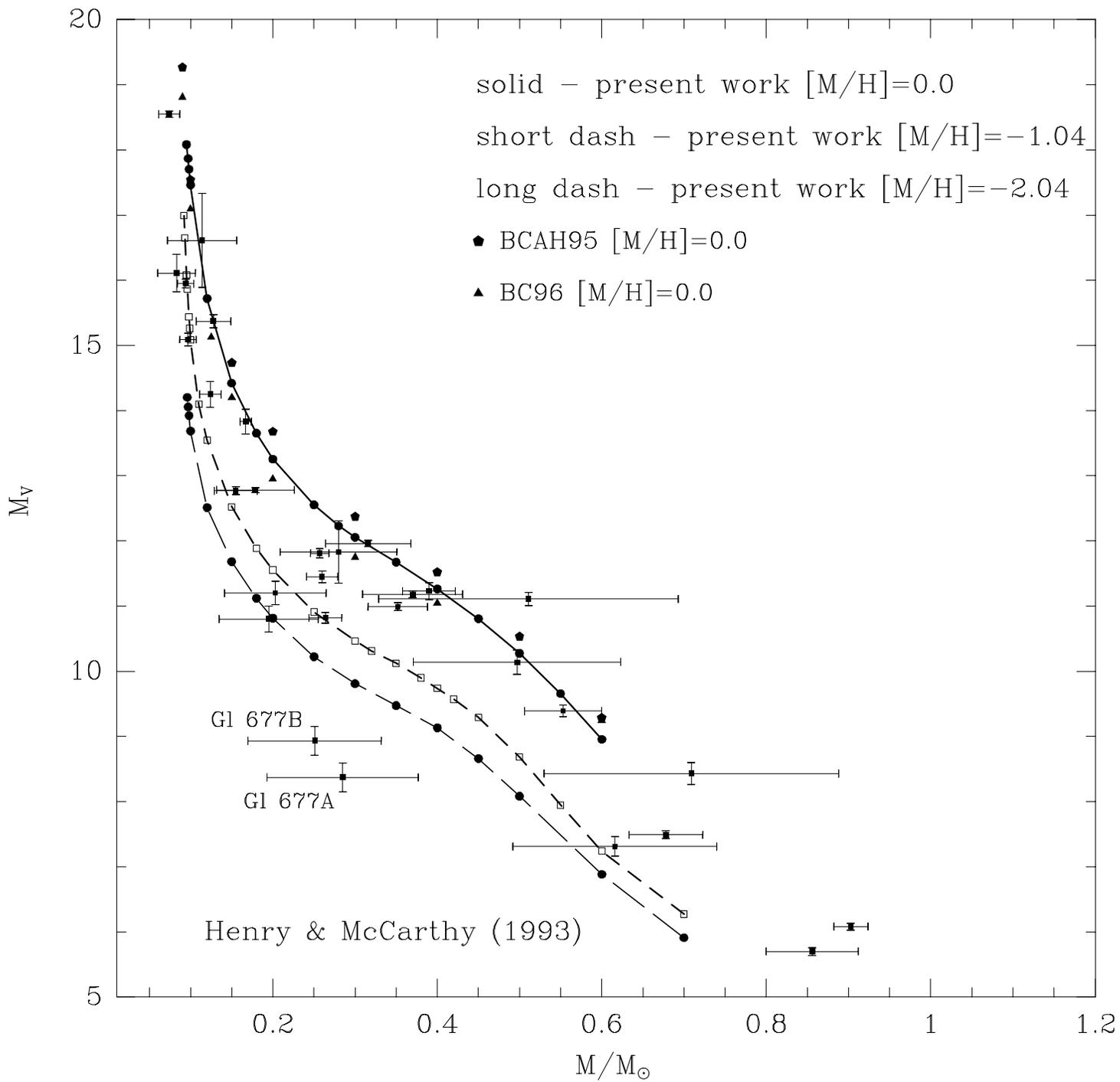
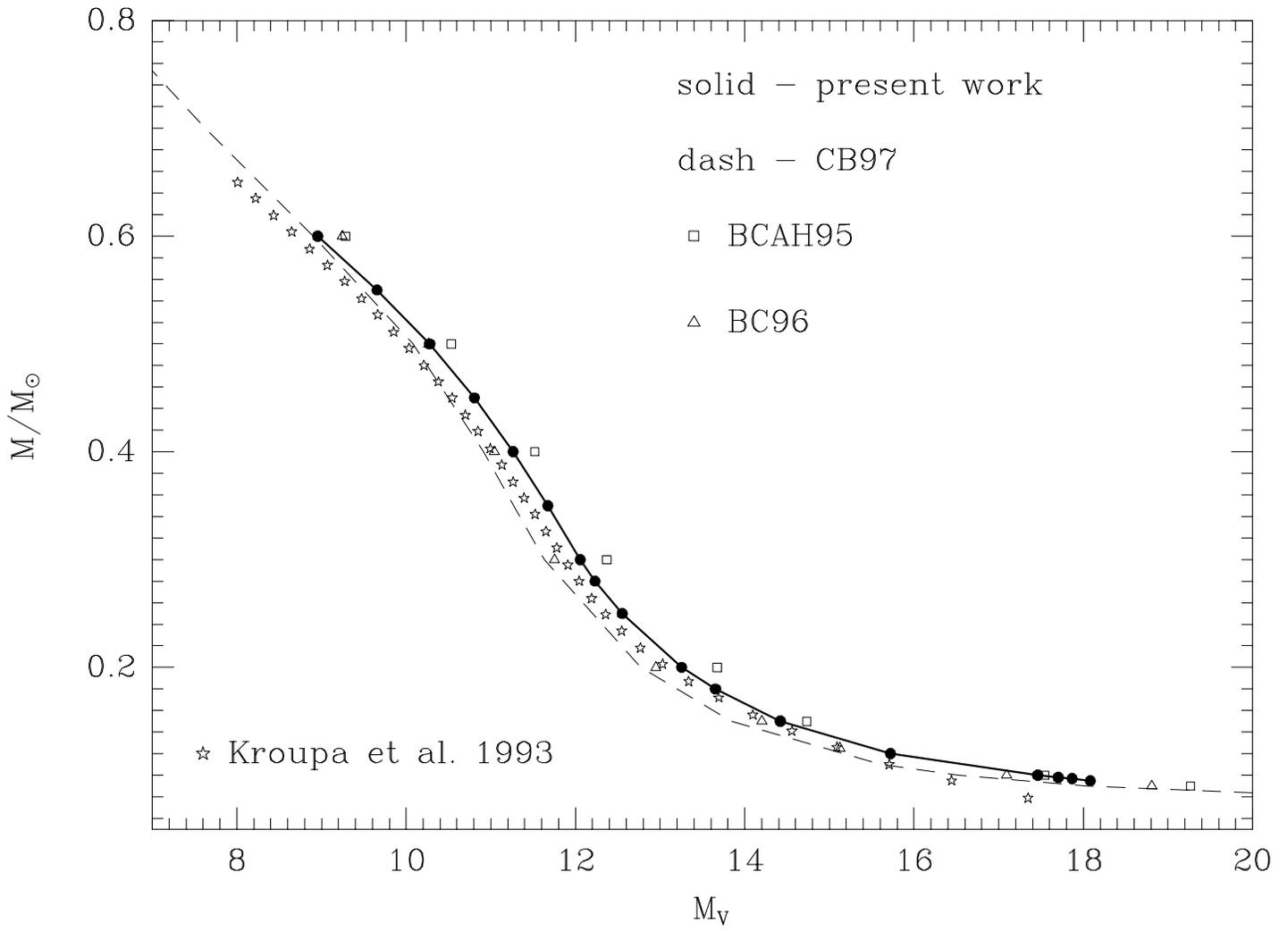
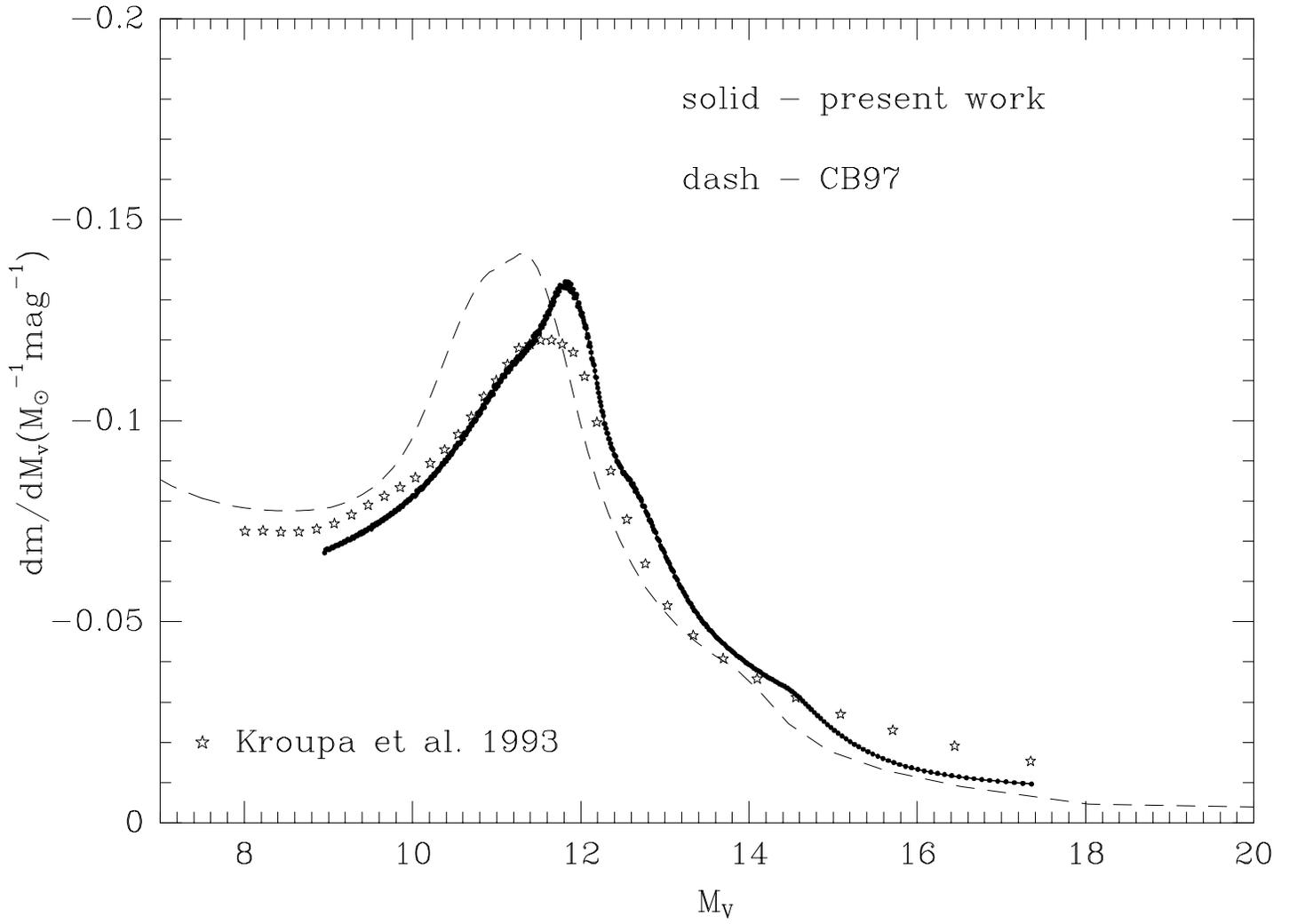
\bye